\documentclass[12pt]{article}
\pdfoutput=1
\usepackage{jheppub}
\usepackage{ucs}
\usepackage[utf8x]{inputenc}
\usepackage{amsmath}
\usepackage{amsfonts}
\usepackage{amssymb}
\usepackage{amsthm}
\usepackage{graphicx}
\usepackage{subfigure}
\usepackage[american]{babel}
\usepackage{fontenc}
\usepackage{graphicx}
\usepackage{mathtools}
\usepackage{float}
\usepackage{bm}  %
\usepackage{ulem}  %
\usepackage[sans]{dsfont}  %

\title{
Zero-Range Effective Field Theory \\
for Resonant Wino  Dark Matter\\  II. Coulomb Resummation}
\author{Eric Braaten,}
\author{Evan Johnson,}
\author{and Hong Zhang}

\affiliation{Department of Physics,
         The Ohio State University, Columbus, OH\ 43210, USA}

\emailAdd{braaten.1@osu.edu}
\emailAdd{johnson.6036@osu.edu}
\emailAdd{zhang.5676@osu.edu}

\abstract{
Near a critical value of the wino mass where there is a zero-energy S-wave resonance at the neutral-wino-pair threshold, low-energy winos can be described by a zero-range effective field theory (ZREFT) in which the winos interact nonperturbatively through a contact interaction and charged winos also have electromagnetic interactions. At energies near the wino-pair thresholds, the Coulomb interaction from photon exchange between charged winos must also be treated nonperturbatively. The parameters of ZREFT can be determined by matching wino-wino scattering amplitudes calculated by solving the Schr\"odinger equation for winos interacting through a potential due to the exchange of electroweak gauge bosons. With Coulomb resummation, ZREFT at leading order gives a good description of the low-energy two-body observables for winos.
}

\smallskip

\begin{document}
\maketitle
\flushbottom

\section{Introduction}
\label{sec:Introduction}

A weakly interacting massive particle ({\it wimp}) is one of the best motivated candidates for a dark-matter particle that provides most of the mass of the universe. A stable particle with weak interactions and whose mass is roughly at the electroweak scale is naturally produced in the early universe with a relic abundance comparable to the observed mass density of dark matter \cite{Kolb:1990vq,Steigman:2012nb}. If the wimp mass $M$ is in the TeV range, the  self-interactions of nonrelativistic wimps are complicated by a nonperturbative effect pointed out by Hisano et al. \cite{Hisano:2002fk}. Weak interactions between low-energy wimps are nonperturbative in the same sense as Coulomb interactions between low-energy charged particles: the exchange of gauge bosons must be summed to all orders in the gauge coupling constant. There can be critical values of the wimp mass where there is a resonance at the wimp-pair threshold. If the wimp mass is near such a critical mass, the annihilation rate of pairs of wimps into electroweak gauge bosons can be enhanced by orders of magnitude  \cite{Hisano:2003ec,Hisano:2004ds}. Wimp-wimp cross sections at low relative velocity can also be increased by orders of magnitude, which can affect the relic abundance of dark matter \cite{Hisano:2006nn,Cirelli:2007xd}.

A resonance in an S-wave channel can generally produce a more dramatic enhancement over a broader range of $M$ than a resonance in a channel with higher orbital angular momentum. There is also a qualitative difference between a near-threshold resonance in an S-wave channel and in a channel for a higher partial wave. The S-wave resonance generates dynamically a length scale that is much larger than the  range of the interactions. This length scale is the absolute value of the S-wave scattering length $a$, which can be orders of magnitude larger than the range. If there are no pair-annihilation channels, the scattering length can even be infinitely large.

In a fundamental quantum field theory, wimps interact through the exchange of electroweak gauge bosons to which they couple through local gauge interactions. The enhancement of low-energy wimp-wimp cross sections can be calculated by summing an infinite set of diagrams in that quantum field theory. The enhancement can be calculated more simply using a nonrelativistic effective field theory (NREFT) in which the wimps have instantaneous interactions at a distance through a potential generated by the exchange of the electroweak gauge bosons. In NREFT, few-body reaction rates of nonrelativistic wimps can be calculated by the numerical solution of a Schr\"odinger equation \cite{Hisano:2002fk}. A thorough development of NREFT for nearly degenerate neutralinos and charginos in the MSSM has been presented in ref.~\cite{Beneke:2012tg}. NREFT has recently been used to calculate the capture rates of two neutral winos into wino-pair bound states through the radiation of a photon \cite{Asadi:2016ybp}.

In the case of an S-wave resonance near threshold, low-energy wimps can be described more simply using a zero-range effective field theory (ZREFT) in which the weak interactions are replaced by zero-range interactions. ZREFT exploits the large length scale that is generated dynamically  by an S-wave resonance. The S-wave scattering length $a$ diverges at critical values of the wimp mass $M$. ZREFT is applicable if $M$ is close enough to a critical value that $|a|$  is  large compared to the range $1/m_W$ of the weak interactions.  ZREFT can be used to calculate analytically wimp-wimp cross sections for wimps  with relative momentum less than $m_W$. There have been several previous applications of zero-range effective field theories to dark matter with resonant S-wave self-interactions. Braaten and Hammer pointed out that the elastic scattering cross section of the dark-matter particles, their total annihilation cross section, and the binding energy and width of a dark matter bound state are all determined by the complex S-wave scattering length \cite{Braaten:2013tza}. Laha and Braaten studied the nuclear recoil energy spectrum in dark-matter direct detection experiments due to both elastic scattering and breakup scattering of an incident dark-matter bound state \cite{Laha:2013gva}. Laha extended that analysis to the angular recoil spectrum in directional detection experiments \cite{Laha:2015yoa}.

In Ref.~\cite{Braaten:2017gpq}, we developed the ZREFT for wimps that consist of the neutral dark-matter particle $w^0$ and charged wimps $w^+$ and $w^-$ with a slightly larger mass. We refer to these wimps as {\it winos}, because the fundamental theory describing them could be the minimal supersymmetric standard model (MSSM) in a region of parameter space where the neutral wino is the lightest supersymmetric particle. The ZREFT for winos  can be organized into a systematically improvable effective field theory by expanding around a renormalization group fixed point. At the RG fixed point, the mass splitting between charged winos and neutral winos is zero, the electromagnetic interactions are turned off, the S-wave unitarity bound is saturated in a scattering channel that is a linear combination of $w^0 w^0$ and $w^+ w^-$, and there is no scattering in the orthogonal channel. In Ref.~\cite{Braaten:2017gpq}, we calculated the wino-wino cross sections analytically in ZREFT without electromagnetism at leading order (LO) and at next-to-leading order (NLO) in the ZREFT power counting. The interaction parameters of ZREFT at LO and at NLO were determined by matching numerical results for scattering amplitudes obtained by solving the Schr\"odinger equation for NREFT. ZREFT at LO  gives fairly accurate predictions for the wino-wino cross sections in the wino-pair threshold region, with the exception of the charged-wino elastic cross section. ZREFT at NLO  gives systematically improved predictions for all the wino-wino cross sections. The power of ZREFT was demonstrated in Ref.~\cite{Braaten:2017gpq} by using it to calculate the formation rate of a wino-pair bound state in the scattering of two neutral winos by a double radiative transition in which two photons are emitted.

In this paper, we extend the results in Ref.~\cite{Braaten:2017gpq} by carrying out the Coulomb resummation of diagrams in which photons are exchanged between pairs of charged winos. We calculate the wino-wino cross sections analytically in ZREFT at LO. The interaction parameters of ZREFT at LO are determined by matching scattering amplitudes with numerical results  obtained by solving the Schr\"odinger equation for NREFT. We show that ZREFT at LO  gives good predictions for the wino-wino cross sections in the wino-pair threshold region. In particular, it reproduces the resonances in the neutral-wino elastic cross section just below the charged-wino-pair threshold and the dramatic oscillations in the charged-wino elastic cross section just above the threshold.

This paper is organized as follows. We begin in section~\ref{sec:winoFTs} by summarizing various quantum field theories that can be used to describe nonrelativistic winos, including the fundamental theory, NREFT, and ZREFT. In section~\ref{sec:NREFT}, we  use the Schr\"odinger equation of NREFT to numerically calculate wino-wino cross sections. In section~\ref{sec:ZRMC}, we calculate wino-wino cross sections with Coulomb resummation analytically in a  field theory with zero-range interactions called the Zero-Range Model. In section~\ref{sec:ZREFTC}, we present analytic results for low-energy two-body observables in ZREFT at LO with Coulomb resummation. We determine the parameters of ZREFT at LO  by matching scattering amplitudes from NREFT. We compare the resulting predictions of ZREFT at LO for  wino-wino cross sections and for the binding energy of a wino-pair bound state with numerical results from solving the Schr\"odinger equation for NREFT. Our results are summarized in section~\ref{sec:Conclusion}. In an Appendix, we solve the Lippmann-Schwinger equations for short-distance transition amplitudes in the Zero-Range Model with Coulomb resummation.

\section{Field Theories for Nonrelativistic Winos}
\label{sec:winoFTs}

In this Section, we summarize field theories that can be used to describe nonrelativistic winos, including the fundamental theory and the effective field theories NREFT and ZREFT.

\subsection{Fundamental theory}
\label{sec:QFT}

We assume the dark-matter particle is the neutral member of an $SU(2)$ triplet of Majorana fermions with zero hypercharge. The Lorentz-invariant quantum field theory that provides a fundamental  description of these fermions could simply be an extension of the Standard Model with this additional $SU(2)$ multiplet and with a symmetry that forbids the decay of the fermion into Standard Model particles. The fundamental theory could also be the Minimal Supersymmetric Standard Model (MSSM) in a region of parameter space where the lightest supersymmetric particle is a wino-like neutralino. In either case, we refer to  the particles in the $SU(2)$ multiplet as {\it winos}. We denote the neutral wino by $w^0$ and the charged winos by $w^+$ and $w^-$.

The relic density of the neutral wino is compatible with the observed mass density of dark matter if the neutral wino mass $M$  is roughly at the electroweak scale \cite{Kolb:1990vq}. We are particularly interested in a mass $M$ at the TeV scale so that effects from the exchange of electroweak gauge bosons between nonrelativistic winos must be summed to all orders. For the neutral wino to be stable, the charged wino must have a larger mass $M+\delta$. In the MSSM, the mass splitting $\delta$ arises from radiative corrections. The splitting from one-loop radiative corrections is determined by $M$ and Standard Model parameters only  \cite{Cheng:1998hc,Feng:1999fu,Gherghetta:1999sw}. As $M$ ranges from 1~TeV to 10~TeV, the one-loop splitting $\delta$ remains very close to 174~MeV. The two-loop radiative corrections decrease $\delta$ by a few MeV  \cite{Ibe:2012sx}. We take the wino mass splitting to be $\delta= 170$~MeV.

The winos can be represented by a triplet $\chi_i$ of 4-component Majorana spinor fields, where the neutral-wino field is $\chi_3$ and the charged-wino fields are linear combinations of $\chi_1$ and $\chi_2$. The most important interactions of the winos are those with the electroweak gauge bosons: the photon, the $W^\pm$, and the $Z^0$. The Lagrangian for the winos is 
\begin{equation}
\mathcal{L}_{\rm wino} = 
\sum_i \left( \tfrac{i}{2} \chi_i^T C \gamma^\mu D_\mu \chi_i  
 - \tfrac{1}{2} M \chi_i^T C \chi_i   \right),
\label{eq:Lwino}
\end{equation}
where $D_\mu$ is the $SU(2)$ gauge-covariant derivative and $C$ is a charge conjugation matrix. The mass $M$ of the winos is an adjustable parameter. The splitting $\delta$ between the masses of $w^\pm$ and $w^0$ arises from electroweak radiative corrections. The relevant Standard Model parameters are the mass $m_W = 80.4$~GeV of the $W^\pm$, the mass $m_Z = 91.2$~GeV of the $Z^0$, the $SU(2)$ coupling constant $\alpha_2= 1/29.5$, the electromagnetic coupling constant $\alpha = 1/137.04$, and the weak mixing angle, which is given by $\sin^2 \theta_w = 0.231$.

\begin{figure}[t]
\centering
\includegraphics[width=0.98\linewidth]{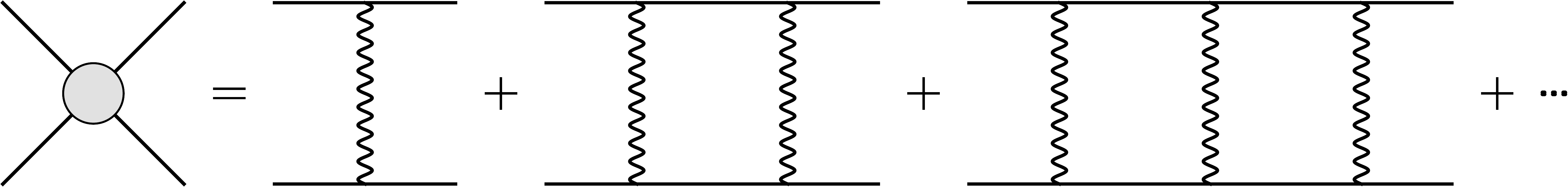}
\caption{Feynman diagrams in the fundamental theory for wino-wino scattering through the exchange of electroweak gauge bosons. The solid lines are neutral winos or charged winos, and the wavy lines are electroweak gauge bosons. If the winos are nonrelativistic, these ladder diagrams must be summed to all orders.}
\label{fig:FundamentalSum}
\end{figure}

Hisano, Matsumoto, and Nojiri pointed out that if the mass of the wino is large enough that $\alpha_2 M$ is of order $m_W$ or larger, loop diagrams in which electroweak gauge bosons are exchanged between nonrelativistic winos are not suppressed \cite{Hisano:2002fk}. The electroweak interactions between a pair of nonrelativistic winos must therefore be treated nonperturbatively by summing ladder  diagrams from the exchange of electroweak bosons between the winos to all orders. For wino-wino scattering, the first few diagrams in the sum are shown in figure~\ref{fig:FundamentalSum}. The resummation of the ladder diagrams to all orders can be carried out more easily by solving a Schr\"odinger equation in a nonrelativistic effective field theory for the winos.

\subsection{Nonrelativistic  effective field theory}
\label{sec:NonrelativisticEFT}

Low-energy winos can be described  by a nonrelativistic effective field theory in which they interact through potentials that arise from the exchange of weak gauge bosons and in which charged winos also have electromagnetic interactions. We call this effective field theory {\it NREFT}. In NREFT, the nonrelativistic wino fields are 2-component spinor fields: $\zeta$ which annihilates a neutral wino $w^0$, $\eta$ which annihilates a charged wino $w^-$, and $\xi$ which creates a charged wino $w^+$. The kinetic terms for winos in the Lagrangian are
\begin{equation}
\mathcal{L}_{\rm kinetic} = \zeta^\dagger\left(i\partial_0+\frac{\bm{\nabla}^2}{2M}\right) \zeta 
+ \eta^\dagger\left(iD_0+\frac{\bm{D}^2}{2M} -\delta\right) \eta
+ \xi^\dagger\left(iD_0-\frac{\bm{D}^2}{2M} +\delta\right) \xi,
\label{eq:kineticLHMN}
\end{equation}
where $D_0$ and $\bm{D}$ are electromagnetic covariant derivatives acting on the charged wino fields. The neutral and charged winos have the same kinetic mass $M$, and the wino mass splitting $\delta$ is taken into account through the rest energy of the charged winos. The weak interaction terms in the Hamiltonian are instantaneous interactions at a distance through a potential produced by the exchange of the $W^\pm$ and $Z^0$ gauge bosons:
\begin{eqnarray}
H_{\rm weak} &=& - \frac12 \int\!\! d^3x \int\!\! d^3y
\bigg(   \frac{\alpha_2 \cos^2 \theta_W}{|\bm{x}-\bm{y}|} e^{-m_Z|\bm{x}-\bm{y}|} 
\eta^\dagger(\bm{x}) \xi(\bm{y})\,  \xi^\dagger(\bm{y}) \eta(\bm{x})~~
\nonumber\\
&&+ \frac{\alpha_2}{|\bm{x}-\bm{y}|}  e^{-m_W|\bm{x}-\bm{y}|} 
\left[ \zeta^\dagger(\bm{x}) \zeta^c(\bm{y})\,  \xi^\dagger(\bm{y}) \eta(\bm{x})
+ \zeta^{c\dagger}(\bm{x}) \zeta(\bm{y})\,  \eta^\dagger(\bm{y}) \xi(\bm{x})\right] \bigg),
\label{eq:LintHMN}
\end{eqnarray}
where $\zeta^c = -i \sigma_2 \xi^*$ and $\sigma_2$ is a Pauli matrix. The potentials from the exchange of $W^\pm$ and $Z^0$ have ranges of order $1/m_W$. The amplitudes for wino-wino scattering can be represented diagrammatically by the same sum of ladder diagrams as in figure~\ref{fig:FundamentalSum}, except that  the wavy  lines for the weak bosons $W^\pm$ and $Z^0$ should be interpreted as instantaneous interactions at a distance through the potentials in eq.~\eqref{eq:LintHMN}.

In refs.~\cite{Hisano:2003ec,Hisano:2004ds}, Hisano, Matsumoto, and Nojiri calculated the  nonperturbative effect of the exchange of electroweak gauge bosons between winos on the annihilation rate of a pair of winos into electroweak gauge bosons by solving a Schr\"odinger equation that can be derived from NREFT. A particularly dramatic consequence is the existence of a zero-energy resonance at the neutral-wino-pair threshold $2M$ at a sequence of critical values of $M$. Near these resonances, the annihilation rate of a wino pair into a pair of electroweak gauge bosons is increased by orders of magnitude. For $\delta = 170$~MeV, the first such resonance is an S-wave resonance at $M=2.39$~TeV.

There are many important momentum scales for nonrelativistic winos. The inverse range of the weak interactions is $m_W = 80.4$~GeV. The momentum scale below which weak interactions are nonperturbative is $\alpha_2 M$. The momentum scale below which electromagnetic interactions are nonperturbative is the  Bohr momentum $\alpha M$. Another important momentum scale is the scale $\sqrt{2M\delta}$ associated with transitions between a neutral-wino pair and a charged-wino pair. For $\delta = 170$~MeV and the first resonance mass $M = 2.39$~TeV, these momentum scales are $\alpha_2 M = 81.1$~GeV, $\alpha M = 17.5$~GeV, and $\sqrt{2M\delta}=28.5$~GeV.

\subsection{Zero-range model}
\label{sec:ZRM}

There can be a resonance at the neutral-wino-pair threshold in any partial wave. An S-wave resonance at the threshold is special, because there is a dynamically generated length scale that is much larger than the range $1/m_W$ of the weak interactions \cite{Braaten:2004rn}. This length scale is the absolute value of the neutral-wino scattering length $a_0$. The corresponding momentum scale $\gamma_0 = 1/a_0$ can be much smaller than any of the other momentum scales provided by interactions described above. For winos with relative momenta small compared to $m_W$, the effects of the exchange of weak bosons can be mimicked by zero-range interactions. Thus winos with sufficiently low energy can be described by a nonrelativistic field theory with local interactions and with electromagnetic interactions. This remains true even if there is an S-wave resonance near the neutral-wino-pair threshold. However in this case, the zero-range interactions must be nonperturbative, because otherwise they cannot generate the large length scale $|a_0|$.

A simple nonrelativistic field theory for low-energy winos with local interactions is the {\it Zero-Range Model} introduced in Ref.~\cite{Braaten:2017gpq}. The winos are described by nonrelativistic two-component spinor fields $w_0$, $w_+$, and $w_-$ that annihilate $w^0$, $w^+$, and $w^-$, respectively. They can be identified with the fields $\zeta$, $\xi^\dagger$, and $\eta$ in NREFT, respectively. The kinetic terms for winos in the Lagrangian for zero-range model are
\begin{equation}
\mathcal{L}_{\rm kinetic} = w_0^\dagger\left(i\partial_0+\frac{\bm{\nabla}^2}{2M}\right) w_0  
+ \sum_\pm w_\pm^\dagger\left(iD_0+\frac{\bm{D}^2}{2M}-\delta\right) w_\pm.  
\label{eq:kineticL}
\end{equation}
The electromagnetic covariant derivatives are
\begin{eqnarray}
D_0 w_\pm = (\partial_0 \pm ieA_0)w_\pm,
\qquad
\bm{D} w_\pm = (\bm{\nabla} \mp ie\bm{A})w_\pm.
\end{eqnarray}
The neutral and charged winos have the same kinetic mass $M$, and the mass splitting $\delta$ is taken into account through the rest energy of the charged winos. Since the neutral wino is a Majorana fermion, a pair of neutral winos can have an S-wave resonance at threshold only in the spin-singlet channel. That channel is coupled to the spin-singlet channel for charged winos. The Lagrangian for zero-range interactions in the spin-singlet channels can be expressed as
\begin{eqnarray}
\mathcal{L}_{\rm zero-range} &=& 
-\tfrac{1}{4} \lambda_{00} ( w_0^{c\dagger} w_0^{d\dagger} )
\tfrac12 ( \delta^{ac}\delta^{bd}- \delta^{ad}\delta^{bc}) ( w_0^a w_0^b )
\nonumber\\
&& 
-\tfrac{1}{2} \lambda_{01} (  w_+^{c\dagger} w_-^{d\dagger} )
\tfrac12 ( \delta^{ac}\delta^{bd}- \delta^{ad}\delta^{bc}) ( w_0^a w_0^b )
\nonumber\\
&& 
-\tfrac{1}{2} \lambda_{01} (  w_0^{c\dagger} w_0^{d\dagger} )
\tfrac12 ( \delta^{ac}\delta^{bd}- \delta^{ad}\delta^{bc}) ( w_+^a w_-^b )
\nonumber\\
&& 
- \lambda_{11} ( w_+^{c\dagger} w_-^{d\dagger} )
\tfrac12 ( \delta^{ac}\delta^{bd}- \delta^{ad}\delta^{bc})
( w_+^a w_-^b ),
\label{eq:ZRint}
\end{eqnarray}
where $\lambda_{00}$, $\lambda_{01}$, and $\lambda_{11}$ are bare coupling constants. The factor $\frac12( \delta^{ac}\delta^{bd}- \delta^{ad}\delta^{bc})$ is the projector onto the spin-singlet channel.

\begin{figure}[t]
\centering
\includegraphics[width=0.98\linewidth]{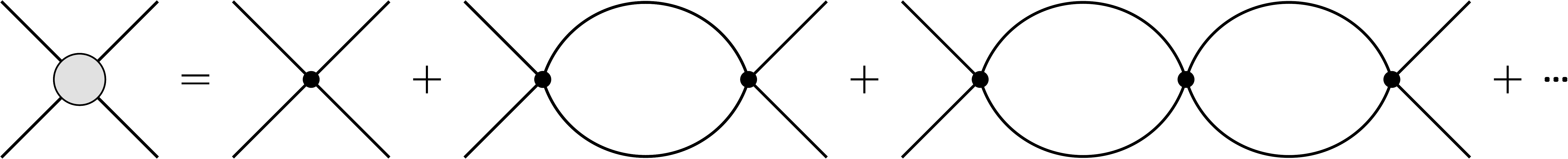}
\caption{Feynman diagrams for wino-wino scattering in the Zero-Range Model without electromagnetism. The solid lines are neutral winos or charged winos. The bubble diagrams  must be summed to all orders. In the Zero-Range Model with electromagnetism, one must also sum ladder diagrams in which photons are exchanged between incoming $w^+$ and $w^-$ lines, between outgoing $w^+$ and $w^-$ lines, and between the $w^+$ and $w^-$ lines in each bubble.}
\label{fig:ZREFTSum0}
\end{figure}

In the Zero-Range Model, the zero-range interactions must be treated  nonperturbatively by summing bubble diagrams involving the vertices from the interaction term in eq.~\eqref{eq:ZRint} to all orders. For wino-wino scattering with $\alpha = 0$, the first few terms in the sum are shown in figure~\ref{fig:ZREFTSum0}. In the Zero-Range Model with electromagnetism, the electromagnetic interactions must also be treated nonperturbatively by summing to all orders ladder diagrams in which photons are exchanged between charged winos. In the absence of electromagnetic interactions, the Zero-Range Model is nonperturbatively renormalizable, at least in the two-wino sector. With electromagnetic interactions included, the Zero-Range Model in Coulomb gauge is probably renormalizable as an effective field theory.

The Zero-Range Model has two coupled scattering channels with different energy thresholds. This model is analogous to the leading order (LO) approximation to the {\it pion-less effective field theory} ($\pi \!\!\!\slash$EFT) that has been widely used in nuclear physics to describe low-energy nucleons \cite{Kaplan:1998tg,Kaplan:1998we}. In $\pi \!\!\!\slash$EFT at LO,  nucleon pairs have two decoupled S-wave scattering channels (the spin-singlet isospin-triplet channel and the spin-triplet isospin-singlet channel) with the same energy threshold. Zero-range models that have two coupled scattering channels with different energy thresholds were first considered in Ref.~\cite{Cohen:2004kf}. They have been applied previously to ultracold atoms \cite{Braaten:2007nq}, to charm meson pairs \cite{Braaten:2007ft}, and to {nucleon-nucleus interactions \cite{Lensky:2011he}.

Coulomb resummation in a zero-range model was first carried out by Kong and Ravndal for the proton-proton system in pion-less effective field theory \cite{Kong:1998sx,Kong:1999sf}. The Coulomb resummation for the two-nucleon system was recently revisited  in Ref.~\cite{Konig:2015aka}, where it was also applied to the three-nucleon system. Coulomb resummation has also been carried out in a zero-range model with two coupled scattering channels with different energy thresholds by Lensky and Birse \cite{Lensky:2011he}.

\subsection{Zero-range effective field theory}
\label{sec:ZREFT}

Range corrections can be incorporated into the Zero-Range Model by adding terms to the Lagrangian with more and  more gradients acting on the fields. Alternatively, for S-wave interactions, range corrections can be incorporated by adding terms to the Lagrangian with more and  more time derivatives acting on the fields \cite{Birse:1998dk}. If all possible range corrections are included, the theory has infinitely many parameters. An {\it effective field theory} can be  defined as a sequence of models with an increasing finite number of parameters that take into account corrections with systematically improving accuracy. A nonrelativistic effective field theory called {\it ZREFT} for winos that have an S-wave resonance near the neutral-wino-pair threshold was introduced in Ref.~\cite{Braaten:2017gpq}. The winos interact through zero-range self-interactions and through their couplings to the electromagnetic field.

An effective field theory can be defined most rigorously through deformations of a renormalization-group (RG) fixed point. Systematically improving accuracy is ensured by adding to the Lagrangian operators with increasingly higher scaling dimensions. In Ref.~\cite{Lensky:2011he}, Lensky and Birse carried out a careful RG analysis of the two-particle sector for a nonrelativistic field theory for distinguishable particles with two coupled scattering channels and with zero-range S-wave interactions. They identified three distinct RG fixed points. The first RG fixed point is the {\it noninteracting fixed point} at which the $2\times 2$ T-matrix is zero at all energies $E$: $\bm{\mathcal{T}}_*(E) = 0$. The second RG fixed point is the {\it two-channel-unitarity fixed point}, in which the cross sections saturate the S-wave unitarity bounds in both scattering channels. At this fixed point, the two scattering channels have the same threshold at $E=0$ and the T-matrix with the standard normalization of states in a nonrelativistic field theory is
\begin{equation}
\label{eq:Tfp2}
\bm{\mathcal{T}}_*(E) = \frac{4\pi i}{M \sqrt{ME}} \,
\begin{pmatrix} 1~  & ~0\\ 0~ & ~1 \end{pmatrix} ,
\end{equation}
where $M$ is the mass of the particle. The cross sections have the scaling behavior $1/E$. The power-law dependence on $E$ reflects the scale invariance of the interactions. In ref.~\cite{Lensky:2011he}, Lensky and Birse pointed out that there is a third RG fixed point: the {\it single-channel-unitarity fixed point}. At this fixed point, the two scattering channels have the same threshold at $E=0$ and the T-matrix is 
\begin{equation}
\label{eq:Tfp3}
\bm{\mathcal{T}}_*(E) = \frac{4\pi i}{M \sqrt{ME}}
\begin{pmatrix} \cos^2\phi  & \cos\phi  \sin\phi \\ 
 \cos\phi  \sin\phi & \sin^2\phi \end{pmatrix}.
\end{equation}
There is nontrivial scattering in a single channel that is a linear combination of the two scattering channels with mixing angle $\phi$. In that channel, the cross section saturates the S-wave unitarity bound. There is no scattering in the orthogonal channel. The single-channel-unitarity fixed point is the most natural one for describing a system with a single fine tuning, such as the tuning of the wino mass $M$  to a unitarity value where there is an S-wave resonance at the threshold.

In ref.~\cite{Lensky:2011he}, Lensky and Birse diagonalized the RG flow near the single-channel-unitarity fixed point whose T-matrix $\bm{\mathcal{T}}_*(E)$ is given  in eq.~\eqref{eq:Tfp3}, identifying all the scaling perturbations and their scaling dimensions. The scaling perturbations provide a basis for the vector space of perturbations near the fixed point. Their coefficients provide a complete parametrization of the T-matrix. There is one relevant scaling perturbation that corresponds to changing $\sqrt{ME}$ in the denominator in eq.~\eqref{eq:Tfp3} to  $\sqrt{ME} + i \gamma$, where $\gamma$ is a real parameter that can be interpreted as an inverse scattering length. There are two marginal scaling perturbations. One of them corresponds to turning on the splitting $2\delta$ between the thresholds in the two channels, and the other corresponds to changing the mixing angle $\phi$. All the other scaling perturbations are irrelevant. The inclusion of scaling perturbations with increasingly higher scaling dimensions defines the successive improvements of ZREFT. The parameters in ZREFT at leading order  (LO) are $M$, $\delta$, the mixing angle $\phi$, and the parameter $\gamma$. There are two additional parameters in ZREFT at next-to-leading order (NLO), and there is one additional parameter at next-to-next-to-leading order  (NNLO).

The systematic improvement provided by the effective field theory can be formulated in terms of an expansion in powers of the ratio of the generic momentum scale $Q$ described by the effective field theory and the smallest momentum scale $\Lambda$ beyond its domain of applicability. In the case of winos, $\Lambda$ can be identified with $m_W$. We take the energy $E$ and the mass splitting $\delta$ to be order $Q^2/M$. The natural scale for $\gamma$ is $\Lambda$, but we assume it is reduced to order $Q$ by the fine tuning responsible for the S-wave resonance near the threshold. The mixing angle scales as $(Q/\Lambda)^0$. All other parameters scale as negative powers of $\Lambda$. Instead of taking the interaction parameters to be coefficients in the Lagrangian, it is convenient to take them to be parameters in the inverse of the $2 \times 2$ T-matrix. The systematically improving accuracy of the effective field theory is ensured by including parameters whose leading contributions to the T-matrix scale with increasingly higher powers of $Q/\Lambda$.

ZREFT can  be extended to an effective field theory for winos and photons. In ZREFT at LO, the only electromagnetic coupling is that of the charged winos through the covariant derivatives acting on the charged wino fields in eq.~\eqref{eq:kineticL}. Thus including electromagnetism does not introduce any additional adjustable parameters at LO. In ZREFT beyond LO, gauge invariance requires some  of the terms proportional to powers of  $E$ in the inverse of the T-matrix to be accompanied by additional interaction terms involving the time component $A_0$ of the photon field. They do not introduce any additional parameters. There may also be additional interaction terms involving the gauge-invariant electromagnetic field strengths $\bm{E}$ and $\bm{B}$, which would introduce additional parameters.

\section{NREFT}
\label{sec:NREFT}

In this section, we use NREFT to calculate cross sections for nonrelativistic wino-wino scattering. We keep the wino mass splitting fixed at $\delta= 170$~MeV, and we study the dependence of the two-wino observables on the wino mass $M$. We also consider the effect of turning off the electromagnetic coupling constant $\alpha$.

\subsection{Schr\"odinger equation}
\label{sec:SchrEq}

Ladder diagrams from the exchange of electroweak gauge bosons between a pair of wimps can be summed to all orders in NREFT by solving a Schr\"odinger equation. The coupled-channel radial Schr\"odinger equation for S-wave scattering in the spin-singlet channel is
\begin{equation}
\left[ -\frac{1}{M} \begin{pmatrix} 1~  & ~0 \\ 0~ & ~1 \end{pmatrix}\left( \frac{d\ }{dr} \right)^2
+ 2\delta \begin{pmatrix} 0~  & ~0 \\ 0~ & ~1 \end{pmatrix}
+\bm{V}(r) \right] r \binom{R_0(r)}{R_1(r)} = E\,  r \binom{R_0(r)}{R_1(r)},
\label{eq:radialSchrEq}
\end{equation}
where $R_0(r)$ and $R_1(r)$ are the radial wavefunctions for a  pair of neutral winos and a pair of charged winos, respectively. The  $2 \times 2$ matrix of potentials is
\begin{equation}
\bm{V}(r) = -  
\alpha_2 \begin{pmatrix}                0               & \sqrt{2}\, e^{-m_Wr}/r \\ 
                         \sqrt{2}\, e^{-m_Wr}/r  &  c_w^2\, e^{-m_Zr}/r   \end{pmatrix}
-  \alpha \begin{pmatrix} 0~  & ~0 \\ 0~ & ~1/r \end{pmatrix},
\label{eq:V-matrix}
\end{equation}
where $c_w = \cos \theta_w$. There is a continuum of positive energy eigenvalues $E$ that correspond to S-wave scattering states. There may also be discrete negative eigenvalues that correspond to S-wave bound states.

The coupled-channel radial Schr\"odinger equation in eq.~\eqref{eq:radialSchrEq} can be solved for the radial wavefunctions $R_0(r)$ and $R_1(r)$. For energy $E$ above the charged-wino-pair threshold $2 \delta$, the asymptotic solutions for $R_0(r)$ and $R_1(r)$ as $r \to \infty$ determine a dimensionless, unitary, and symmetric $2 \times 2$ S-matrix $\bm{S}(E)$. The dimensionless $2\times2$ T-matrix $\bm{T}(E)$ is defined by
\begin{equation}
\bm{S}(E) =
\mathds{1} + i \, \bm{T}(E),
\label{eq:S-high}
\end{equation}
where $\mathds{1}$ is the $2\times 2$ unit matrix. The T-matrix satisfies the unitarity equation
\begin{equation}
2 \, \textrm{Im}\, \bm{T}(E) = \bm{T} ^\dagger(E)\, \bm{T}(E) .
\label{eq:T-unitarity}
\end{equation}
For energy in the range $0<E<2\delta$, the asymptotic solutions for $R_0(r)$ determine a $1 \times 1$ S-matrix whose single element can be expressed as $S_{00}(E) = \exp\big(2 i \delta_0(E)\big)$, where $\delta_0(E)$ is the real-valued S-wave phase shift.

We denote the contribution to the cross section for elastic scattering from channel $i$ to channel $j$ at energy $E$ from scattering in the S-wave spin-singlet channel by $\sigma_{i \to j}(E)$. The expressions for these cross sections in terms of the T-matrix elements $T_{ji}$ are
\begin{subequations}
\begin{eqnarray}
\sigma_{0 \to j}(E) &=& \frac{2\pi}{M^2 v_0(E)^2} \big| T_{j0}(E) \big|^2,
\label{eq:sig0j-T}
\\
\sigma_{1\to j}(E) &=&   \frac{\pi}{M^2 v_1(E)^2}\big| T_{j1}(E) \big|^2,
\label{eq:sig1j-T}
\end{eqnarray}
\label{eq:sigij-T}%
\end{subequations}
where $v_0(E)$ and $v_1(E)$ are the wino velocities in the center-of-mass frame for a neutral-wino pair and a charged-wino pair with total energy $E$:
\begin{subequations}
\label{eq:v0,1-E}
\begin{eqnarray}
v_0(E) &=&  \sqrt{E/M},
\label{eq:v0-E}
 \\
v_1(E) &=&  \sqrt{(E-2 \delta)/M}.
\label{eq:v1-E}
\end{eqnarray}
\end{subequations}
For the neutral-wino elastic cross section $\sigma_{0 \to 0}$, the energy threshold is $E=0$. For the other three cross sections $\sigma_{1 \to 0}$, $\sigma_{0 \to 1}$,  and $\sigma_{1 \to 1}$, the energy threshold is $E=2\delta$. The cross sections in eqs.~\eqref{eq:sigij-T} have been averaged over initial spins and summed over final spins. The {\it S-wave unitarity bounds} for the scattering of $w^0w^0$, which are identical spin-$\tfrac12$ particles, and for the scattering of $w^+w^-$, which are distinguishable spin-$\tfrac12$ particles, are
\begin{subequations}
\begin{eqnarray}
\sigma_{0\to 0}(E)  &\le& \frac{8 \pi}{ME},
\label{sigma-unitarity0}
\\
\sigma_{1\to 1}(E)  &\le& \frac{4 \pi}{M(E-2\delta)}.
\label{sigma-unitarity1}
\end{eqnarray}
\label{sigma-unitarity}
\end{subequations}

\subsection{Neutral-wino elastic scattering}
\label{sec:w0Scattering}

\begin{figure}[t]
\centering
\includegraphics[width=0.8\linewidth]{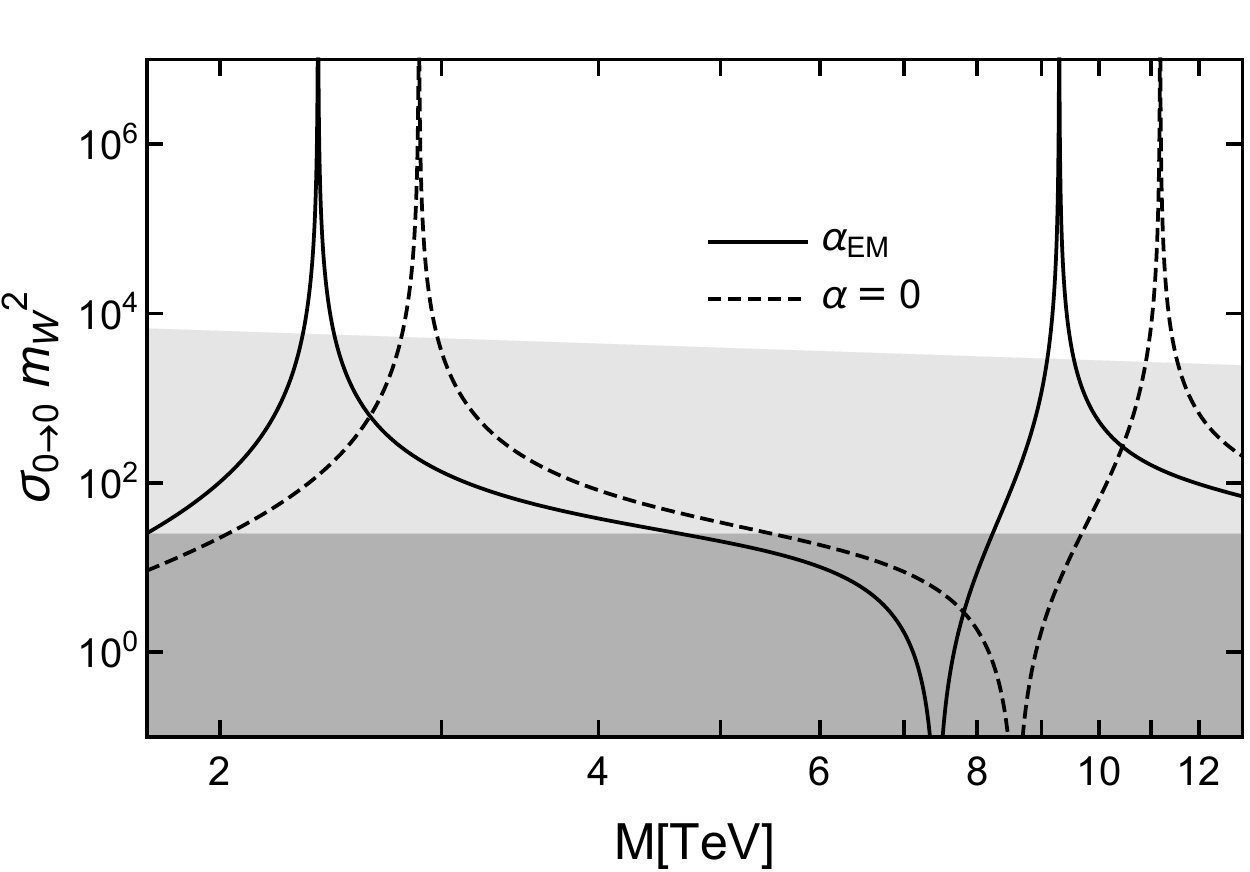}
\caption{Neutral-wino elastic cross section $\sigma_{0 \to 0}$ at zero energy as a function of the wino mass $M$. The cross section is shown for $\alpha=1/137$  (solid curve) and for $\alpha = 0$ (dashed curve). The darker shaded region is $\sigma_{0 \to 0} < 8 \pi/m_W^2$ and the lighter shaded region is $\sigma_{0 \to 0} < 4 \pi/M \delta$. If $\sigma_{0 \to 0}$ is above the  darker shaded region, the ZREFT for neutral and charged winos is applicable. If $\sigma_{0 \to 0}$ is above the lighter shaded region, a ZREFT for neutral winos only is applicable.}
\label{fig:sigma00vsMcode}
\end{figure}

The neutral-wino elastic cross section $\sigma_{0 \to 0}(E=0)$ at zero energy for $\delta=170$~MeV is shown as a function of the wino mass $M$ in figure~\ref{fig:sigma00vsMcode}. The cross section diverges at critical values of $M$. The first critical mass is $M_*=2.39$~TeV and the second is 9.23~TeV. The divergence indicates that there is a zero-energy resonance at the neutral-wino-pair threshold. At a critical mass where there is an S-wave resonance at the neutral-wino-pair threshold, the neutral-wino elastic cross section saturates the unitarity bound in eq.~\eqref{sigma-unitarity0} in the limit $E\to 0$. We therefore refer to such a critical mass as a {\it unitarity mass}, and we refer to a system with such a mass as being {\it at unitarity}.

The neutral-wino elastic cross section at zero energy depends sensitively on the strength $\alpha$ of the Coulomb potential. The Coulomb potential can be  turned off by setting $\alpha = 0$ in the potential matrix in eq.~\eqref{eq:V-matrix}. The resulting cross section for neutral winos with zero energy is compared to the cross section at the physical value $\alpha= 1/137$ in figure~\ref{fig:sigma00vsMcode}. If the Coulomb potential is turned off by setting $\alpha = 0$, the cross section at $M=2.39$~TeV is reduced to $123/m_W^2$. The shape of the curve is almost the same, but the first two unitarity masses are shifted  upward by about 20\% to 2.88~TeV and 11.18~TeV. 

\begin{figure}[t]
\centering
\includegraphics[width=0.8\linewidth]{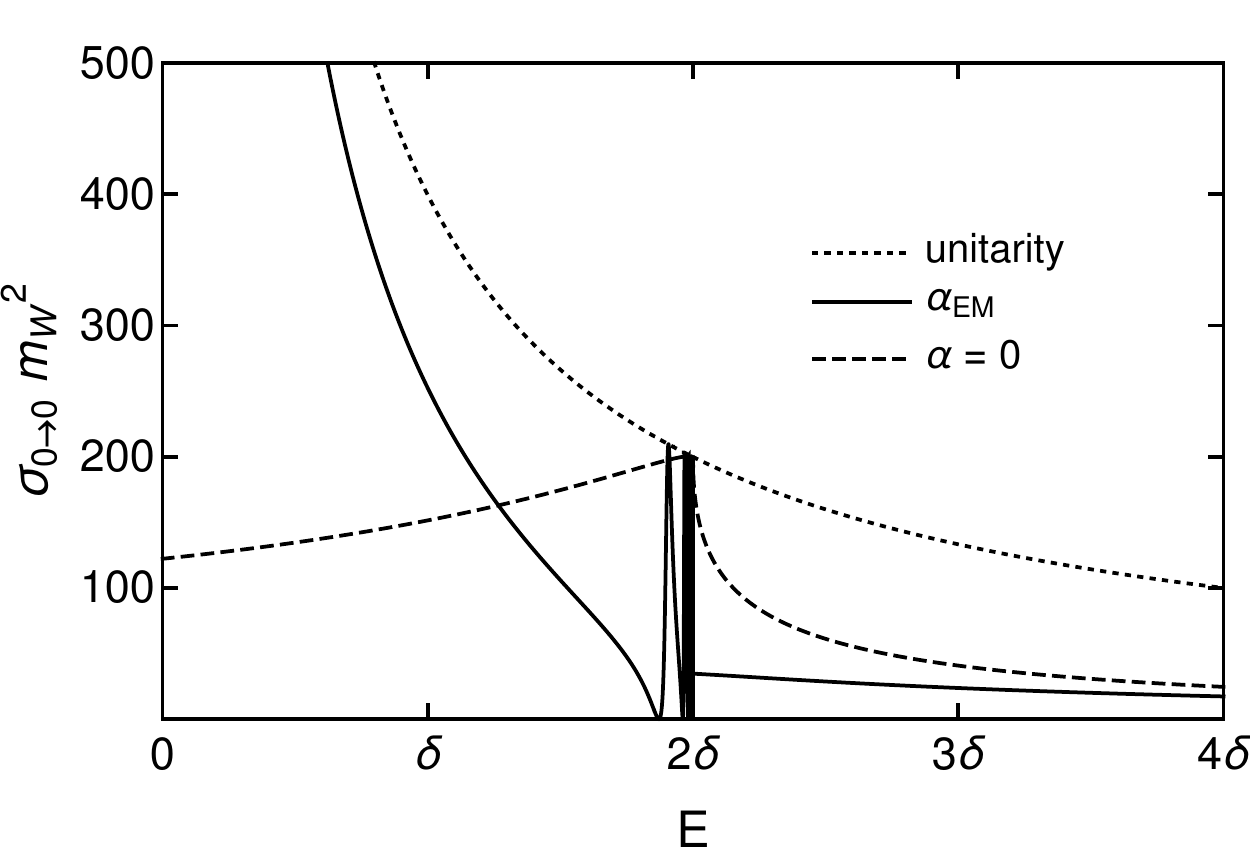}
\caption{Neutral-wino elastic cross section $\sigma_{0 \to 0}$  as a function of the energy $E$. The cross section for $M_*=2.39$~TeV is shown for $\alpha = 1/137$ (solid curve) and for $\alpha = 0$ (dashed curve). The S-wave unitarity bound is shown as a dotted curve.}
\label{fig:sigma00-NREFT}
\end{figure}

The neutral-wino elastic cross section $\sigma_{0 \to 0}(E)$ has the most dramatic energy dependence at a unitarity mass, such as $M_*=2.39$~TeV. The cross section for $M_*=2.39$~TeV is shown in figure~\ref{fig:sigma00-NREFT}. As $E$ approaches 0, the cross section approaches the unitarity bound in eq.~\eqref{sigma-unitarity0} from below, saturating the bound in the limit. Just below the charged-wino-pair threshold $2 \delta$, the cross section with the Coulomb potential has a sequence of narrow resonances whose peaks saturate the unitarity bound. The resonances can be interpreted as bound states in the Coulomb potential for the charged-wino pair $w^+w^-$. Just above the threshold at $2\delta$, the cross section is $34.7/m_W^2$, and it decreases slowly as $E$ increases. The cross section with $\alpha =0$ and $M=2.39$~TeV is also shown in figure~\ref{fig:sigma00-NREFT}, and it has a qualitatively different behavior. As $E$ approaches 0, the cross section has a finite limit. The resonances just below the charged-wino-pair threshold disappear. As $E$ increases from 0, the cross section increases monotonically until the threshold $2 \delta$, where it has a kink, and it then decreases as $E$ increases further.

Neutral winos with energies well below the charged-wino-pair threshold $2\delta$ have short-range interactions, because the Coulomb interaction enters only through virtual charged winos. The short-range interactions guarantee that $v_0(E)/T_{00}(E)$ can be expanded in powers of the relative momentum $p = \sqrt{ME}$:
\begin{equation}
\frac{2M\,v_0(E)}{T_{00}(E)} =
-\frac{1}{a_0}  - i p+ \frac12 r_0\,p^2 + \frac18 s_0\,p^4 + {\cal O}(p^6).
\label{eq:T00NRinv}
\end{equation}
The only odd power of $p$ in the expansion is the pure imaginary term $-ip$. The coefficients of the even powers of $p$ are real valued. The leading term in the expansion defines the {\it neutral-wino S-wave scattering length} $a_0$. It diverges at a unitarity mass. The coefficients of $p^2$ and $p^4$ define the {\it effective range}  $r_0$ and a {\it shape parameter} $s_0$.

\begin{figure}[t]
\centering
\includegraphics[width=0.8\linewidth]{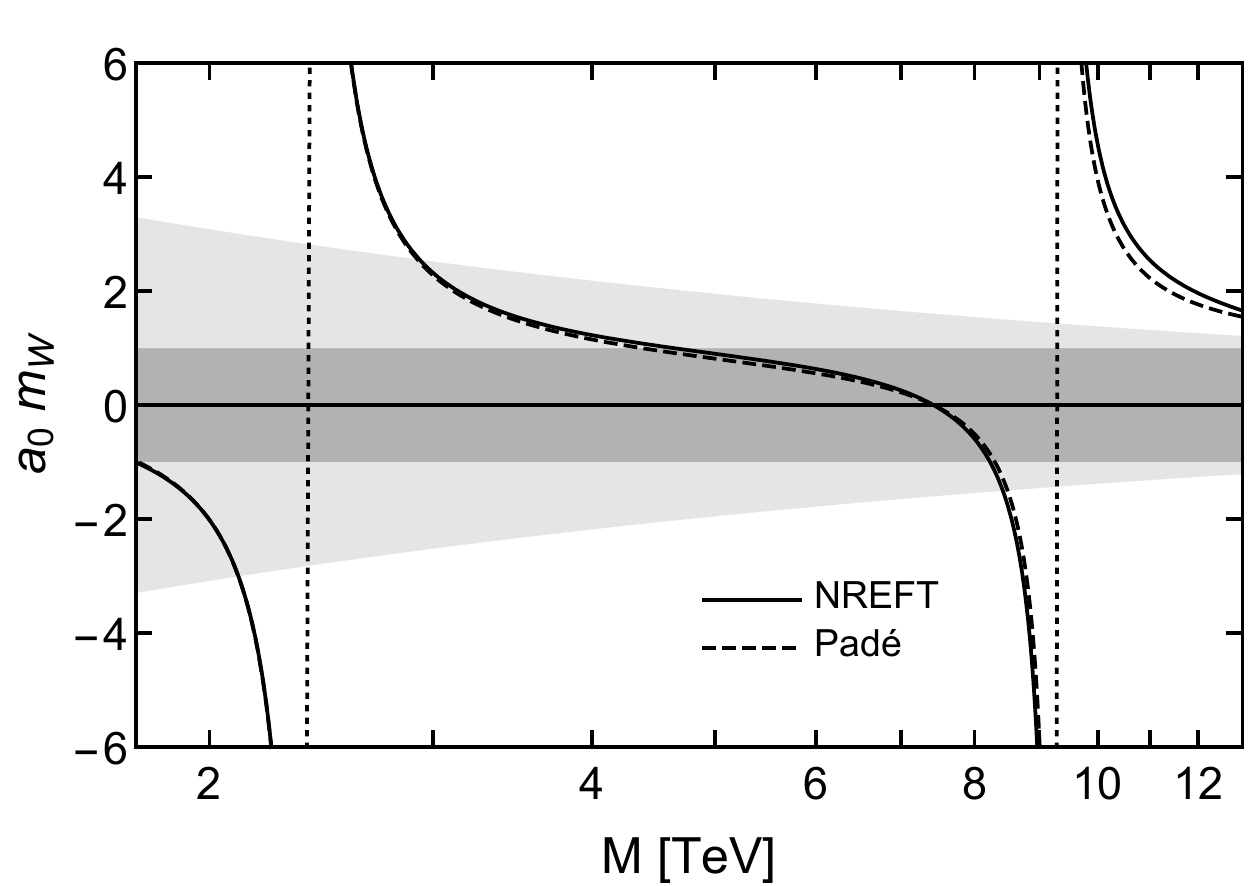}
\caption{Neutral-wino scattering length $a_0$ as a function of the wino mass $M$ (solid curve). The dashed curve is the Pad\'e approximant  given in eq.~\eqref{eq:a0Pade}. The vertical dotted lines indicate the first and second unitarity masses $M_* = 2.39$~TeV and 9.23~TeV. The darker shaded region is $|a_0| <1/m_W$ and the lighter shaded region is $|a_0| <1/\sqrt{2 M \delta}$. If $a_0$ is outside the  darker shaded region, the ZREFT for neutral and charged winos is applicable. If $a_0$ is outside the lighter shaded region, a ZREFT for neutral winos only is applicable.}
\label{fig:Real_a_vsM}
\end{figure}

The coefficients in the range expansion in eq.~\eqref{eq:T00NRinv} can be determined numerically by solving the Schr\"odinger equation. The scattering length $a_0$ for $\delta = 170$~MeV is shown as a function of the wino mass $M$ in figure~\ref{fig:Real_a_vsM}. The dependence of $a_0$ on $M$ can be fit surprisingly well by a Pad\'e approximant in $M$ of order [2,2] whose poles match the first and second resonances at $M_*=2.39$~TeV and  $M_*' = 9.23$~TeV and whose zeros match the first and second zero crossings at $M_0 = 0.0027$~TeV and $M_0' = 7.39$~TeV. The only adjustable parameter is an overall prefactor. We can improve the fit near the resonance at $M_*$ significantly by fitting $M_0$ as well as the prefactor. The resulting fit is
\begin{equation}
\label{eq:a0Pade}
a_0(M) = \frac{0.952}{m_W} \, \frac{(M-M_0)(M-M_0')}{(M-M_*)(M-M_*')},
\end{equation}
where $M_0 = 0.845$~TeV. Near a unitarity mass where $a_0$ diverges, the scattering length is necessarily very sensitive to $\alpha$. If $\alpha$ is set to 0, the scattering length at $M=2.39$~TeV is reduced to $-2.21/m_W$.

The winos can be described by a zero-range effective field theory (ZREFT) for neutral and charged winos if the neutral-wino scattering length is large compared to the range of the weak interactions: $|a_0| > 1/m_W$. Figure~\ref{fig:Real_a_vsM} shows that the region of $M$ near $M_* = 2.39$~TeV in which the 2-channel ZREFT is applicable is roughly from 1.8~TeV to 4.6~TeV. The energy region in which it is applicable is total wino-pair energy $E$ below about $m_W^2/M$, which at $M_*$ is about 2700~MeV. There is a narrower range of $M$ is which neutral winos can be described by a ZREFT for neutral winos only. The neutral-wino scattering length must be large not only compared to $1/m_W$ but also compared to the range associated with the transition between a neutral-wino pair and a virtual  charged-wino pair: $|a_0| > 1/\Delta$, where $\Delta = (2 M \delta)^{1/2}$. Figure~\ref{fig:Real_a_vsM} shows that the region of $M$ in which the ZREFT for neutral winos only is applicable is roughly from 2.1~TeV to 2.9~TeV. The energy region in which it is applicable is total neutral-wino-pair energy $E$ below about $\delta=170$~MeV.

The coefficients of terms with higher powers of $p^2$ in the range expansion in eq.~\eqref{eq:T00NRinv} can also be determined numerically by solving the Schr\"odinger equation. For $\delta = 170$~MeV and $\alpha = 1/137$, the effective range and the shape parameter at the unitarity mass $M_* = 2.39$~TeV are
\begin{subequations}
\begin{eqnarray}
r_0(M_*) &=& -1.653/\Delta_*,
\label{eq:r0*EM}
\\
s_0(M_*) &=& - 2.653/\Delta_*^3,
\label{eq:s0*EM}
\end{eqnarray}
\label{eq:r0,s0*EM}%
\end{subequations}
where  $\Delta_*  = \sqrt{2 M_* \delta} = 28.5$~GeV. The absolute values of the coefficients are order 1, indicating that $\Delta_*$ is an appropriate momentum scale. If the Coulomb potential is turned off by setting $\alpha = 0$, the coefficients on the right sides of eqs.~\eqref{eq:r0*EM} and \eqref{eq:s0*EM} are changed to $-1.224$ and $- 1.878$. Thus these coefficients are somewhat sensitive to $\alpha$.

\begin{figure}[t]
\centering
\includegraphics[width=0.8\linewidth]{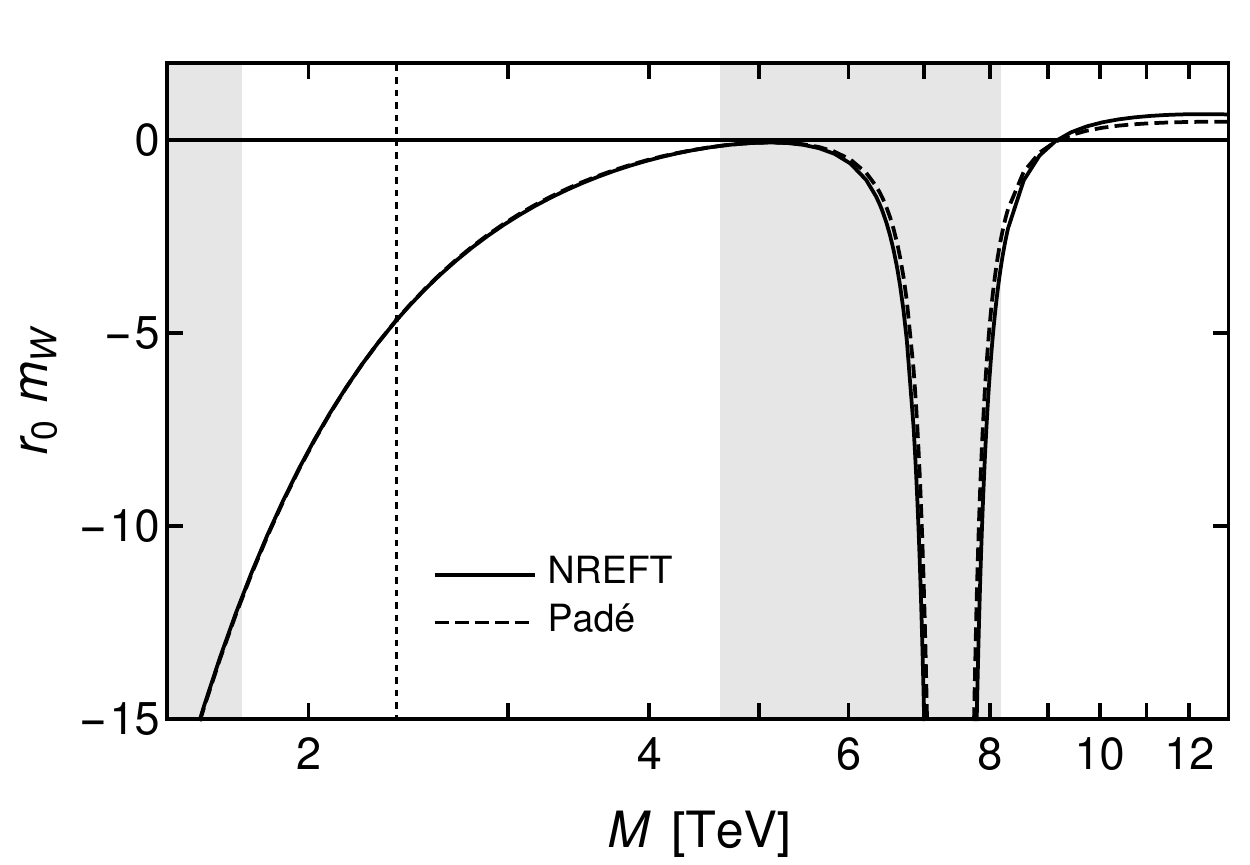}
\caption{Neutral-wino effective range $r_0$ as a function of the wino mass $M$ (solid curve). The dashed curve is the Pad\'e approximant given in eq.~\eqref{eq:r0Pade}. The vertical dotted line indicates the first unitarity mass $M_* = 2.39$~TeV. The grey region is the range of $M$ in which $|a_0| < 1/m_W$, so a ZREFT for neutral and charged winos is not applicable.}
\label{fig:r0vsM}
\end{figure}

The effective range $r_0$ for $\delta = 170$~MeV is shown as a function of the mass $M$ in figure~\ref{fig:r0vsM}. The dependence of $r_0$ on $M$ can be fit surprisingly well by a [4,4] Pad\'e approximant in $M$.  If an offset equal to the local maximum near $M' = 5.13$~TeV is subtracted, the remainder can be fit by a [3,4] Pad\'e with double poles at the zero crossings $M_0$ and $M_0'$ of $a_0(M)$, a double zero at $M' = 5.13$~TeV, and a single zero at $M'' = 9.11$~TeV. The only adjustable parameter in the Pad\'e approximant is an overall prefactor. We choose to improve the fit near the resonance at $M_*$ by fitting $M_0$ as well as the prefactor. The resulting Pad\'e approximant is
\begin{equation}
\label{eq:r0Pade}
r_0(M) =  (5.063/m_W) \left( \frac{M_*(M-M')^2 (M-M'')}{(M-M_0)^2(M-M_0')^2} - 0.0109 \right),
\end{equation}
where $M_0 = 0.124$~TeV.

Particles with short-range interactions that produce an S-wave resonance sufficiently close to their scattering threshold have universal low-energy behavior that is completely determined by their S-wave scattering length $a_0$ \cite{Braaten:2004rn}. The universal predictions are just those of the  single-channel ZREFT at leading order. The universal approximation to the cross section is
\begin{equation}
\sigma_{0 \to 0}(E) = \frac{8\pi}{1/a_0^2 +  ME}.
\label{eq:sigma00-largea}
\end{equation}
The universal region is where $|a_0|$ is large compared to the range set by the interactions. For neutral winos, the appropriate range is the maximum of $1/m_W$ and  $1/\Delta$. The universal region of $M$ is inside the region from 2.1~TeV to 2.9~TeV. The universal approximation becomes increasingly accurate as $M$ approaches the unitarity mass $M_*=2.39$~TeV. The universal region of the energy is $E \ll 2\delta$.

\subsection{Charged-wino scattering}
\label{sec:w+/-Scattering}

\begin{figure}[t]
\centering
\includegraphics[width=0.493\linewidth]{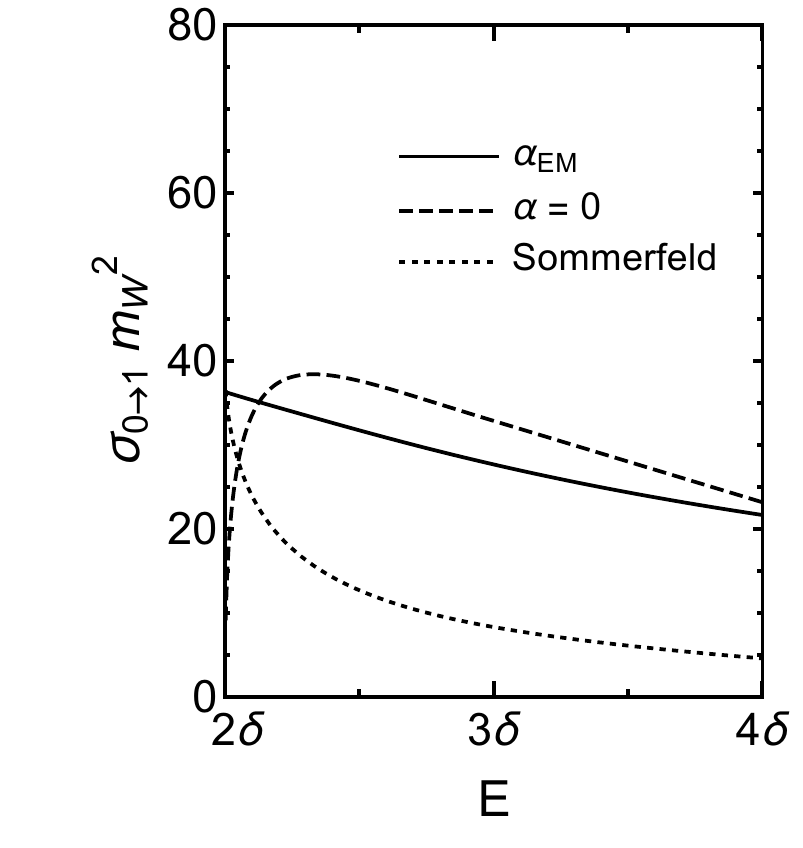}
~
\includegraphics[width=0.46\linewidth]{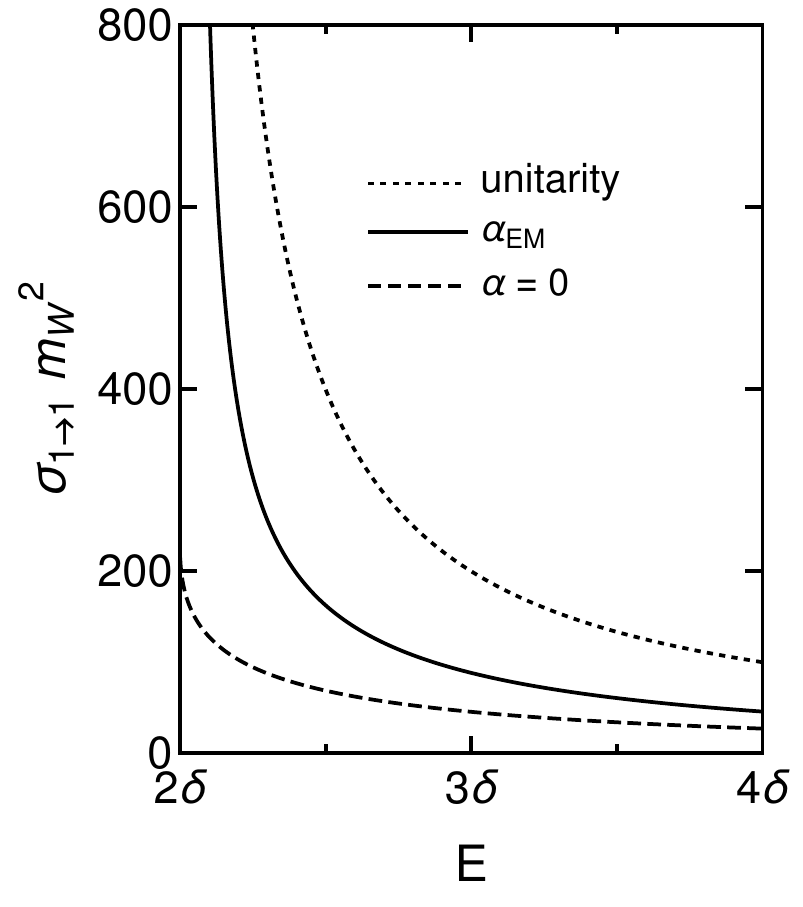}
\caption{Neutral-to-charged transition cross section $\sigma_{0 \to 1}$ (left panel) and the charged-wino elastic cross section $\sigma_{1 \to 1}$ (right panel) as functions of the energy $E$. The cross sections for $M_*=2.39$~TeV are shown for $\alpha = 1/137$ (solid curves)  and for $\alpha = 0$ (dashed curves). In the left panel, the dotted curve is the cross section $\sigma_{0 \to 1}$ for $\alpha = 0$ multiplied by the Sommerfeld factor $C^2(E)$ and normalized to the cross section for $\alpha = 1/137$ at the threshold. In the right panel, the dotted curve is the S-wave unitarity bound in eq.~\eqref{sigma-unitarity1}.}
\label{fig:sigma01,11-NREFT}
\end{figure}

The energy dependence of the neutral-to-charged transition cross section $\sigma_{0 \to 1}(E)$ at the unitarity mass $M_*=2.39$~TeV is illustrated in the left panel of figure~\ref{fig:sigma01,11-NREFT}. As $E$ decreases to the threshold $2 \delta$, the cross section increases monotonically to $36.3/m_W^2$. The cross section with $\alpha =0$ is also shown in the left panel of figure~\ref{fig:sigma01,11-NREFT}, and it has a qualitatively different behavior. As $E$ decreases towards $2 \delta$, the cross section with $\alpha =0$ increases  to a maximum near $E=2.39\,\delta$, and it then decreases to zero. The zero comes from a factor of $v_1(E)$ from the phase space of the final-state $w^+ w^-$. The nonzero cross section at the threshold for $\alpha=1/137$ is due to a Sommerfeld factor for Coulomb rescattering of the final-state $w^+$ and $w^-$. The nonperturbative effect of the Coulomb rescattering of charged particles was  derived by Sommerfeld around 1920 \cite{Sommerfeld:1921}. As the energy $E$ approaches the threshold $2\delta$, the cross section differs from the cross section for $\alpha = 0$  by a multiplicative factor that is the product of a constant and a Sommerfeld factor that depends on the velocity $v_1(E)$ of the charged particles. For particles with charges $\pm1$, the Sommerfeld factor is
\begin{equation}
C^2(E) =
  \frac{\pi \alpha/v_1}{1- \exp(- \pi \alpha/v_1)}.
\label{eq:Sommerfeld0}
\end{equation}
The Sommerfeld factor approaches 1 as $v_1$ increases, and it approaches $\pi \alpha/v_1$ as $v_1 \to 0$. The factor of $1/v_1$ cancels the factor of $v_1$ from the phase space of the final-state $w^+ w^-$, so the cross section has a nonzero limit as $v_1 \to 0$. If the momentum scale $\alpha M$ in the Sommerfeld factor was much smaller than the other relevant momentum scales, the cross section for $\alpha = 1/137$ at relative momentum of order $\alpha M$ could be approximated by the cross section for $\alpha = 0$ multiplied by the Sommerfeld factor $C^2(E)$ and normalized to the cross section for $\alpha = 1/137$ at the threshold. This approximation is shown as a dotted line in the left panel of figure~\ref{fig:sigma01,11-NREFT}. It is not a good approximation, because the momentum scale $\sqrt{2M\delta} = 28.5$~GeV associated with neutral-to-charged transitions is comparable to the momentum scale $\alpha M = 17.5$~GeV in the Sommerfeld factor.

The energy dependence of the charged-wino elastic cross section $\sigma_{1 \to 1}(E)$ at the unitarity mass $M_*=2.39$~TeV is illustrated in the right panel of figure~\ref{fig:sigma01,11-NREFT}. As $E$ decreases to the threshold $2 \delta$, the cross section appears to increase monotonically to infinity. However at energies $E$ extremely close to the threshold at $2 \delta$, there are rapid oscillations in the cross section that are too large to be visible in figure~\ref{fig:sigma01,11-NREFT}. The cross section with $\alpha =0$ is also shown in the right panel of figure~\ref{fig:sigma01,11-NREFT}, and it has a qualitatively different behavior. As $E$ decreases towards  $2 \delta$, it increases monotonically to a finite maximum.

\section{Zero-Range Model  with Coulomb Resummation}
\label{sec:ZRMC}

In this Section, we calculate analytically the transition amplitudes for $w^0 w^0$ and $w^+ w^-$ in the Zero-Range Model with Coulomb resummation.

\subsection{Transition amplitudes}
\label{sec:TransAmps}

Observables in the sector consisting of two neutral winos $w^0w^0$ or two charged winos $w^+ w^-$ are conveniently encoded in the amplitudes for transitions among the two coupled channels. We denote the neutral channel $w^0 w^0$ by the index 0 and the charged channel $w^+ w^-$ by the index 1. In the Zero-Range Model, the S-wave spin-singlet transition amplitudes $\mathcal{A}_{ij}(E)$ are functions of the total energy $E$ of the wino pair only. The T-matrix elements $\mathcal{T}_{ij}(E)$ for  S-wave wino-wino scattering are obtained by evaluating the transition amplitudes $\mathcal{A}_{ij}(E)$ on the energy shell. The constraints on the T-matrix elements from S-wave unitarity can be derived from the unitarity condition for the amplitude matrix $\bm{\mathcal{A}}(E)$ at real $E$, which can be expressed as 
\begin{equation}
\bm{\mathcal{A}}(E)- \bm{\mathcal{A}}(E)^* = 
- \frac{1}{8 \pi} \bm{\mathcal{A}}(E) \bm{M}^{1/2} \Big[ \bm{\kappa}(E)  - \bm{\kappa}(E)^* \Big]  
\bm{M}^{1/2} \bm{\mathcal{A}}(E)^*,
\label{eq:A-unitarity}
\end{equation}
where $\bm{M}$ is the $2\times2$ diagonal matrix
\begin{equation}
\label{eq:Mmatrix}
\bm{M}=
\begin{pmatrix}  M &   0   \\ 
                          0  & 2M   
\end{pmatrix}
\end{equation}
 and $\bm{\kappa}$ is a diagonal matrix whose entries are functions of $E$:
\begin{equation}
\bm{\kappa}(E) =
\begin{pmatrix} \kappa_0(E)  &          0       \\ 
                                 0            & \kappa_1(E)
\end{pmatrix} .
\label{eq:kappamatrix}
\end{equation}
The functions $\kappa_0$ and $\kappa_1$ of the complex energy $E$ have branch points at 0 and $2\delta$, respectively:
\begin{subequations}
\begin{eqnarray}
\kappa_0(E) &=& \sqrt{-ME-i\varepsilon},
\label{eq:kappa0}
\\
\kappa_1(E) &=& \sqrt{-M(E-2\delta)-i\varepsilon}.
\label{eq:kappa1}
\end{eqnarray}
\label{eq:kappa01}%
\end{subequations}
The different diagonal entries of the matrix $\bm{M}$ in eq.~\eqref{eq:Mmatrix} are a convenient way to take into account that the neutral channel $w^0w^0$ consists of a pair of identical fermions while the charged channel $w^+w^-$ consists of two distinguishable fermions.

The amplitudes $\mathcal{A}_{ij}(E)$ have a diagrammatic representation. We represent the propagator for the neutral wino $w^0$ by a solid line without an arrow. We represent the propagator for the charged winos $w^+$ and $w^-$ by solid lines with a forward arrow and a backward arrow, respectively. We represent the photon propagator by a wavy line. The interaction term in the Lagrangian for the Zero-Range Model in eq.~\eqref{eq:ZRint} provides vertices for $w^0w^0 \to w^0w^0$, $w^0w^0 \to w^+w^-$, $w^+w^- \to w^0w^0$, and $w^+w^- \to w^+w^-$. The covariant derivatives in the kinetic term in eq.~\eqref{eq:kineticL} provide vertices in which one or two photons attach to a $w^+$ line or to a $w^-$ line. The transition amplitude $\mathcal{A}_{ij}(E)$ can be expressed  as the sum of all diagrams with the appropriate pair of incoming solid  lines  and the appropriate pair of outgoing  solid  lines.

\subsection{Coulomb amplitude}
\label{sec:CoulombAmp}

\begin{figure}[t]
\centering
\includegraphics[width=0.90\linewidth]{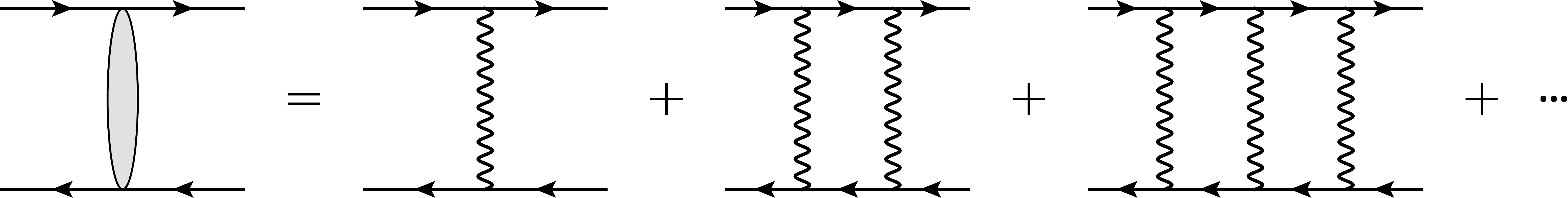}
\caption{Amplitude for $w^+ w^- \to w^+ w^-$ with Coulomb interactions only. The ladder diagrams from the exchange of a photon between $w^+$ and $w^-$ must be summed to all orders.}
\label{fig:Coulomb}
\end{figure}

If the only interactions between winos were the Coulomb interactions between charged winos, the only nonzero transition amplitude in the two-wino sector with zero total charge would be the amplitude for $w^+ w^- \to w^+ w^-$. The amplitude would be given by the sum of ladder diagrams in figure~\ref{fig:Coulomb}. The projection  $\mathcal{A}_{11}$ of that amplitude onto the S-wave channel is the S-wave Coulomb transition amplitude:
\begin{equation}
\label{eq:ACoulomb}
\mathcal{A}_C(E) = 
\left( 1 - \frac{\Gamma(1+i \eta)}{\Gamma(1-i \eta)}  \right) \frac{2\pi}{M\, \kappa_1(E)} ,
\end{equation}
where $\kappa_1$ is given in eq.~\eqref{eq:kappa1} and $\eta$ is an energy variable defined by
\begin{equation}
\label{eq:eta-def}
\eta(E) \equiv 
 \frac{i\, \alpha \, M}{2\, \kappa_1(E)} 
 = i \frac{\alpha M}{2} \big[ - M(E-2 \delta) - i \epsilon \big]^{-1/2}.
\end{equation}
For a real energy $E = 2 \delta + p^2/M$ above the charged-wino-pair threshold, $\eta$ is real and negative: $\eta = - \alpha M/2p$. For a real energy $E$ below the charged-wino-pair threshold, $\eta$ is pure imaginary. The amplitude in eq.~\eqref{eq:ACoulomb} has poles in $E$  at real energies $E_n$ that correspond to Coulomb bound states of $w^+ w^-$:
\begin{equation}
\label{eq:E-n}
E_n = 2 \delta
- \frac{ \alpha^2 M}{4 n^2} ,
\end{equation}
where $n$ is a positive integer. The value of $\eta$ at the energy $E_n$ of a Coulomb bound state is $\eta_n = i n$. For real energy $E$, the Coulomb transition amplitude satisfies the unitarity condition 
\begin{equation}
\label{eq:ACunitarity}
\mathcal{A}_C(E) - \mathcal{A}_C(E)^* = 
- \frac{M}{4\pi} \mathcal{A}_C(E) \big[ \kappa_1(E) - \kappa_1(E)^* \big] \mathcal{A}_C(E)^*.
\end{equation}
For $E < 2 \delta$, $\kappa_1$ is real and $\eta$ is pure imaginary, so the unitarity condition in eq.~\eqref{eq:ACunitarity} is satisfied because both sides vanish. For $E > 2 \delta$, $\kappa_1$ is pure imaginary and $\eta$ is real. The unitarity condition in eq.~\eqref{eq:ACunitarity} then follows from the explicit  expression for the Coulomb transition amplitude in eq.~\eqref{eq:ACoulomb}.

\subsection{Coulomb resummation}
\label{sec:Coulomb}

\begin{figure}[t]
\centering
\includegraphics[width=0.98\linewidth]{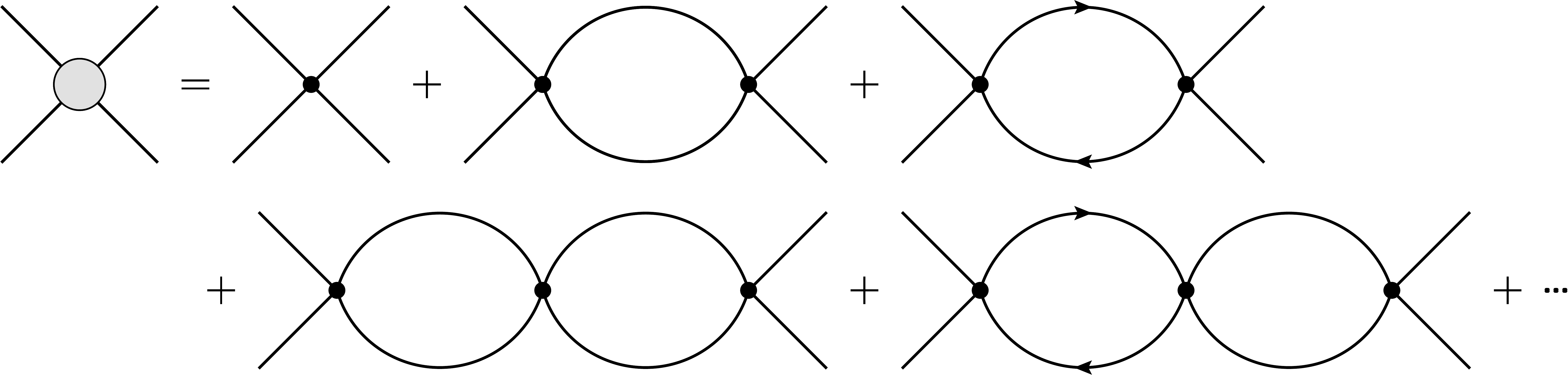}
\caption{Feynman diagrams for the transition amplitude $\mathcal{A}_{00}(E)$ for $w^0 w^0 \to w^0 w^0$ in the Zero-Range Model without electromagnetism. The bubble diagrams must be summed to all orders. Each bubble can be either a neutral-wino pair without arrows or a charged wino pair with arrows.}
\label{fig:ZREFTSum}
\end{figure}

In the absence of electromagnetic interactions, the transition amplitude $\mathcal{A}_{00}(E)$ for $w^0 w^0 \to w^0 w^0$ is given by the sum of all bubble diagrams, as illustrated in figure~\ref{fig:ZREFTSum}. The Feynman diagrams for $w^0 w^0 \to w^+ w^-$, $w^+ w^- \to w^0 w^0$, and $w^+ w^- \to w^+ w^-$ are obtained by putting arrows on the outgoing pair of lines, on the incoming pair of lines, and on both, respectively. The transition amplitudes can be determined analytically by solving Lippmann-Schwinger equations. The Lippmann-Schwinger equations for the Zero-Range Model without Coulomb interactions were solved nonperturbatively in Appendix~A of Ref.~\cite{Braaten:2017gpq}. The solution is expressed most simply by giving the inverse of the $2\times2$ matrix $\bm{\mathcal{A}}(E)$:
\begin{equation}
\label{eq:Ainverse0}
\bm{\mathcal{A}}^{-1}(E) = \frac{1}{8\pi} \bm{M}^{1/2}
\Big[ - \bm{\gamma} + \bm{\kappa}(E) \Big] \bm{M}^{1/2},
\end{equation}
where  $\bm{\gamma}$ is a symmetric matrix of renormalized parameters:
\begin{equation}
\label{eq:gammamatrix}
\bm{\gamma}= 
\begin{pmatrix} \gamma_{00}   & \gamma_{01} \\ 
 \gamma_{01} & \gamma_{11} 
\end{pmatrix}.
\end{equation}
The unitarity equation in eq.~\eqref{eq:A-unitarity} is automatically satisfied if the parameters $\gamma_{00}$, $\gamma_{01}$, and $\gamma_{11}$ are real.

Since charged winos also have electromagnetic interactions, there are additional diagrams for wino-wino scattering beyond those  in figure~\ref{fig:ZREFTSum}. The additional diagrams have photons exchanged between charged wino lines. Most of the diagrams have effects that are suppressed by  one or more factors of the electromagnetic coupling constant $\alpha = 1/137$. However if the relative momentum of a pair of charged winos is of order $\alpha M$ or smaller, there are photon-exchange diagrams that are not suppressed. In Coulomb gauge, the diagrams that are not suppressed are ladder diagrams in which static Coulomb photons are exchanged between a pair of charged winos. The summation of these diagrams is called {\it Coulomb resummation}.

\begin{figure}[t]
\centering
\includegraphics[width=0.98\linewidth]{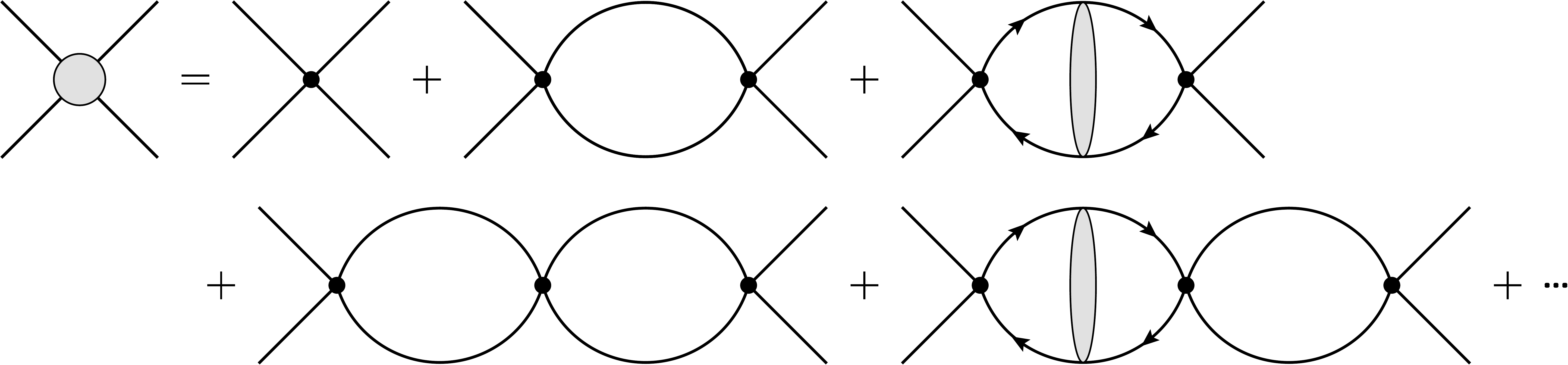}
\caption{Feynman diagrams for the transition amplitude $\mathcal{A}_{00}(E)$ for $w^0 w^0 \to w^0 w^0$ in the Zero-Range Model with Coulomb resummation. The bubble diagrams must be summed to all orders. Each bubble can be either a neutral-wino-pair bubble, which is a one-loop subdiagram, or a charged-wino-pair bubble, which is the sum of the diagrams in figure~\ref{fig:bubbleC}.}
\label{fig:ZREFTSumC}
\end{figure}

\begin{figure}[t]
\centering
\includegraphics[width=0.8\linewidth]{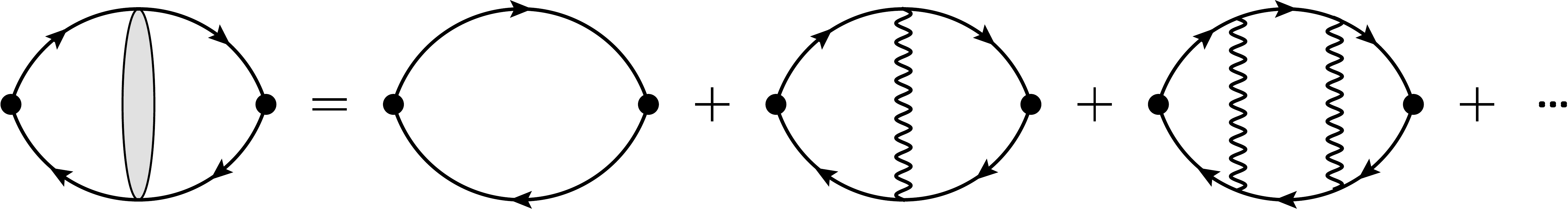}
\caption{Feynman diagrams for the bubble amplitude for $w^+ w^-$ in the Zero-Range Model with Coulomb resummation. The ladder diagrams from the exchange of a photon between $w^+$ and $w^-$ must be summed to all orders.}
\label{fig:bubbleC}
\end{figure}

\begin{figure}[t]
\centering
\includegraphics[width=0.75\linewidth]{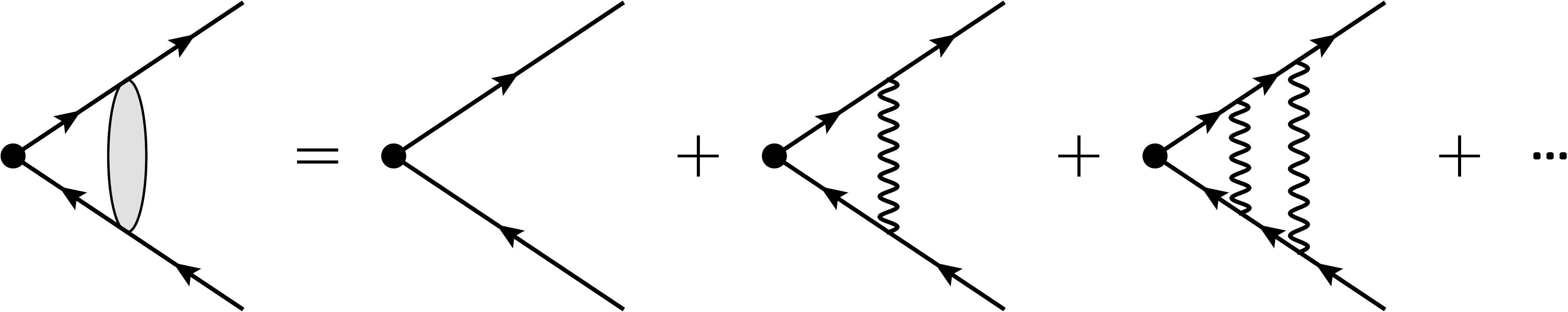}
\caption{Feynman diagrams for the amplitude for creation of $w^+ w^-$ at a point in the Zero-Range Model with Coulomb resummation. The ladder diagrams from the exchange of a photon between $w^+$ and $w^-$ must be summed to all orders. The sum is equal to the tree diagram multiplied by the amplitude $W_1(E)$ in eq.~\eqref{eq:S-eta}.}
\label{fig:pointC}
\end{figure}

For the amplitude for $w^0 w^0 \to w^0 w^0$, Coulomb resummation involves adding all ladder diagrams in which photons are exchanged between the $w^+$ and $w^-$ inside the charged-wino bubbles, as illustrated in figure~\ref{fig:ZREFTSumC}. The charged-wino bubble with Coulomb resummation is the sum of the diagrams in figure~\ref{fig:bubbleC}. For amplitudes with $w^+ w^-$ in the initial state and/or in the final state, Coulomb resummation also  involves adding all ladder diagrams in which photons are exchanged between the incoming  $w^+$ and $w^-$ lines and/or the outgoing $w^+$ and $w^-$ lines. For the outgoing $w^+$ and $w^-$ lines, Coulomb resummation involves replacing the final vertex by the sum of diagrams in figure~\ref{fig:pointC}. Finally, Coulomb resummation for $w^+ w^- \to w^+ w^-$ also requires adding the diagrams in figure~\ref{fig:Coulomb}  in which photons are exchanged between $w^+$ and $w^-$. 

If winos have short-range interactions as well as the Coulomb interactions between charged winos, the matrix of S-wave transition amplitudes can be expressed in the form
\begin{equation}
\label{eq:A-sC}
\bm{\mathcal{A}}(E) = 
\begin{pmatrix} ~0~  & 0  \\  0  & \mathcal{A}_C(E) \end{pmatrix} 
+
\begin{pmatrix} ~1~  & 0  \\  0  & W_1(E) \end{pmatrix} 
\bm{\mathcal{A}}_s(E)
\begin{pmatrix} ~1~  & 0  \\  0  & W_1(E) \end{pmatrix} ,
\end{equation}
where $W_1(E)$ is the dimensionless amplitude for creating or annihilating $w^+$ and $w^-$ with total energy $E$ at a point in the presence of Coulomb interactions, and the matrix $\bm{\mathcal{A}}_s(E)$ is the contribution to $\bm{\mathcal{A}}(E)$ from diagrams in which the first interaction and the last interaction are both short-range interactions. We refer to its entries as the  {\it short-distance transition amplitudes}. The amplitude $W_1(E)$ for creating $w^+w^-$ at a point can be obtained diagrammatically by expressing the sum of diagrams in figure~\ref{fig:pointC} as the tree diagram multiplied by $W_1(E)$. It is determined in Appendix~\ref{app:LSeq}:
\begin{equation}
W_1(E) = C(E) \, \left( \frac{\Gamma(1 + i \eta)}{\Gamma(1 - i \eta)} \right)^{1/2} ,
\label{eq:S-eta}
\end{equation}
where $\eta$ is the function of $E$ in eq.~\eqref{eq:eta-def} and  $C$ is the square root of the Sommerfeld factor in  eq.~\eqref{eq:Sommerfeld0}:
\begin{equation}
C^2(E)  = \frac{2 \pi \eta}{\exp(2 \pi \eta) - 1}.
\label{eq:Sommerfeld}
\end{equation}
In the Zero-Range Model, the short-distance  transition amplitudes $\mathcal{A}_{s,ij}(E)$ can be determined analytically by solving the Lippmann-Schwinger equations in Appendix~\ref{app:LSeq}. The solution is expressed most simply by giving the inverse of the $2\times2$ matrix $\bm{\mathcal{A}}_s(E)$:
\begin{equation}
\label{eq:Ainverse0C}
\bm{\mathcal{A}}_s^{-1}(E) = \frac{1}{8\pi} \bm{M}^{1/2}
\Big[ - \bm{\gamma} + \bm{K}(E) \Big] \bm{M}^{1/2},
\end{equation}
where $\bm{\gamma}$ is the symmetric matrix of renormalized parameters in eq.~\eqref{eq:gammamatrix} and $\bm{K}$ is the diagonal matrix
\begin{equation}
\bm{K}(E) =
\begin{pmatrix} \kappa_0(E)  &          0       \\ 
                                 0            & K_1(E)
\end{pmatrix} .
\label{eq:Kmatrix}
\end{equation}
Its first diagonal entry is the function $\kappa_0$ in eq.~\eqref{eq:kappa0}, and its second diagonal entry is
\begin{equation}
K_1(E) =\alpha M \left[ \psi(i \eta) + \frac{1}{2 i \eta} - \log(-i\eta)  \right],
\label{eq:K1-E}
\end{equation}
where $\psi(z)=(d/dz)\log \Gamma(z)$ and $\eta(E)$ is  defined in eq.~\eqref{eq:eta-def}. This function has a logarithmic branch point at $E=2\delta$ and poles at the Coulomb bound-state energies in eq.~\eqref{eq:E-n}. Since $\eta$ is negative for $E > 2 \delta$, the argument of the logarithm in eq.~\eqref{eq:K1-E} should be interpreted  as $e^{+i\pi }i \eta$. As $z \to \infty$ in any direction of the complex plane except along the negative real axis, the asymptotic behavior of $\psi(z) $ is
\begin{equation}
\psi(z) \longrightarrow \log(z) - \frac{1}{2z} - \frac{1}{12z^2} + \ldots.
\label{eq:psi-largez}
\end{equation}
This implies that $K_1(E)$ approaches a constant as $E$ approaches the threshold $2 \delta$ from above:
\begin{equation}
K_1(2\delta^+) = -i \pi \alpha M .
\label{eq:K1-threshold}
\end{equation}
The function $K_1(E)$ does not have a limit as $E$ approaches $2 \delta$ from below, because $\psi(z)$ has poles at the negative integers.

\section{ZREFT with Coulomb resummation}
\label{sec:ZREFTC}

In this section, we present the transition amplitudes for $w^0 w^0$ and $w^+ w^-$ in ZREFT at LO with Coulomb resummation. We determine the adjustable parameters of ZREFT at LO by matching low-energy $w^0 w^0$ scattering amplitudes from NREFT. We compare predictions of ZREFT at LO for wino-wino cross sections and for the binding energy of a wino-pair bound state with results from NREFT.

\subsection{Transition amplitudes}
\label{sec:AmpsZREFT}

In order to give an explicit parametrization of the transition amplitudes $\mathcal{A}_{ij}(E)$ for ZREFT, we introduce two 2-component unit vectors that depend on the mixing angle $\phi$:
\begin{equation}
\label{eq:u,v-def}
\bm{u}(\phi) = \binom{\cos\phi}{\sin\phi}, \qquad \bm{v}(\phi) = \binom{-\sin\phi}{~~\cos\phi}.
\end{equation}
We use these vectors to define two projection matrices and another symmetric matrix:
\begin{subequations}
\begin{eqnarray}
\bm{\mathcal{P}}_u(\phi) &=& \bm{u}(\phi)\, \bm{u}(\phi)^T
= \begin{pmatrix} \cos^2\phi  & \cos\phi  \sin\phi \\  \cos\phi  \sin\phi & \sin^2\phi \end{pmatrix}, 
\label{eq:Pu}
\\ 
\bm{\mathcal{P}}_v(\phi) &=& \bm{v}(\phi)\, \bm{v}(\phi)^T
= \begin{pmatrix} \sin^2\phi  & - \cos\phi  \sin\phi \\  - \cos\phi  \sin\phi & \cos^2\phi \end{pmatrix},
\label{eq:Pv}
\\ 
\bm{\mathcal{P}}_m(\phi) &=& \bm{u}(\phi)\, \bm{v}(\phi)^T + \bm{v}(\phi)\, \bm{u}(\phi)^T
= \begin{pmatrix} -\sin(2\phi)  & \cos(2\phi) \\  \cos(2\phi) & \sin(2\phi) \end{pmatrix}.
\label{eq:Pm}
\end{eqnarray}
\label{eq:Pu,Pv,Pm}%
\end{subequations}
The superscript $T$ on $\bm{u}$ or $\bm{v}$ indicates the transpose of the column vector. The three matrices defined in eqs.~\eqref{eq:Pu,Pv,Pm} form a basis for $2 \times2$ symmetric matrices. This set of matrices is closed under differentiation:
\begin{subequations}
\begin{eqnarray}
\bm{\mathcal{P}}_u'(\phi) &=& \bm{\mathcal{P}}_m(\phi), 
\label{eq:dPu}
\\ 
\bm{\mathcal{P}}_v'(\phi) &=& -\bm{\mathcal{P}}_m(\phi), 
\label{eq:dPv}
\\ 
\bm{\mathcal{P}}_m'(\phi) &=& -2\, \bm{\mathcal{P}}_u(\phi) +2\, \bm{\mathcal{P}}_v(\phi).
\label{eq:dPm}
\end{eqnarray}
\label{eq:dPu,dPv,dPm}%
\end{subequations}
The T-matrix at the RG fixed point for ZREFT with $\alpha=0$ is
\begin{equation}
\label{eq:Tfp3wino}
\bm{\mathcal{T}}_*(E) = \frac{8\pi i}{\sqrt{ME}} \, \bm{M}^{-1/2} 
\, \bm{\mathcal{P}}_u(\phi)\,  \bm{M}^{-1/2},
\end{equation}
where $\bm{M}$ is the diagonal matrix in eq.~\eqref{eq:Mmatrix}.

A possible choice for the interaction parameters of ZREFT with $\alpha = 0$ are the coefficients of the scaling perturbations to the Lagrangian near the RG  fixed point that corresponds to the T-matrix in eq.~\eqref{eq:Tfp3wino}. A more convenient choice are coefficients in the expansion in powers of $E$ of  the inverse $\bm{\mathcal{T}}^{-1}(E)$ of the T-matrix. The corresponding parameterization for the inverse of the matrix of transition amplitudes is
\begin{eqnarray}
\label{eq:AinverseZREFT}
\bm{\mathcal{A}}^{-1}(E) = \frac{1}{8\pi} \bm{M}^{1/2} 
\Big[ \big(- \gamma_u + \tfrac12 r_u p^2 + \ldots \big) \bm{\mathcal{P}}_u(\phi)
 + \big(-1/a_v + \ldots \big) \bm{\mathcal{P}}_v(\phi)
 \nonumber
 \\
 + \big(\tfrac12 r_m p^2 + \ldots \big) \bm{\mathcal{P}}_m(\phi) + \bm{\kappa}(E) \Big]
\bm{M}^{1/2},
\end{eqnarray}
where $p^2 = ME$ and $\bm{\kappa}(E)$ is the diagonal matrix in eq.~\eqref{eq:kappamatrix}. The coefficients of  $\bm{\mathcal{P}}_u$, $\bm{\mathcal{P}}_v$, and $\bm{\mathcal{P}}_m$ have been expanded in powers of $p^2$.  The mixing angle $\phi$ has been chosen so that the $p^0$ term in the expansion of the coefficient of $\bm{\mathcal{P}}_m$ is 0. The interaction parameters of ZREFT with $\alpha = 0$ are the mixing angle $\phi$, the parameters $\gamma_u$ and $a_v$, and the coefficients of the positive powers of $p^2$, such as $r_u$ and $r_m$. These parameters should all be regarded as functions of $M$ and $\delta$ with expansions in powers of $\Delta^2 = 2 M \delta$. The successive improvements of ZREFT can be obtained by successive truncations of the expansions in $p^2$. At leading order (LO), the only nonzero term in the three expansions is the coefficient $-\gamma_u$ of $\bm{\mathcal{P}}_u(\phi)$. By setting the coefficients of positive powers of $p^2$ to zero in $\bm{\mathcal{A}}^{-1}(E)$ in eq.~\eqref{eq:AinverseZREFT}, inverting the matrix, and then taking the limit $a_v \to 0$, we obtain the matrix of transition amplitudes for ZREFT at LO with $\alpha = 0$:
\begin{equation}
\label{eq:AmatrixLO}
\bm{\mathcal{A}}(E) = 
\frac{8\pi}{-\gamma_u +\cos^2\phi\,  \kappa_0(E)  +\sin^2\phi \, \kappa_1(E)}  \, 
\bm{M}^{-1/2} \, \bm{\mathcal{P}}_u(\phi)\,  \bm{M}^{-1/2}.
\end{equation}
At next-to-leading order (NLO), there are two additional interaction parameters: $a_v$ and  $r_u$. At NNLO, there is one additional interaction parameter:  $r_m$.

If $\alpha$ is not zero, it is necessary to resum the effects of the exchange of Coulomb photons between charged winos to all orders. ZREFT at LO with Coulomb resummation is just a limiting case of the Zero-Range Model. The matrix of transition amplitudes $\bm{\mathcal{A}}(E)$ has the form in eq.~\eqref{eq:A-sC}, where $\mathcal{A}_C(E)$ is the Coulomb amplitude in eq.~\eqref{eq:ACoulomb}, $W_1(E)$ is the amplitude for creating $w^+ w^-$ at a point in eq.~\eqref{eq:S-eta}, and $\bm{\mathcal{A}}_s(E)$ is the matrix of short-distance transition amplitudes, which can be expressed as:
\begin{equation}
\label{eq:AsmatrixLOC}
\bm{\mathcal{A}}_s(E) =  \lim_{a_v \to 0} 8\pi  \, \bm{M}^{-1/2}
\big[ - \gamma_u  \bm{\mathcal{P}}_u(\phi) -(1/a_v) \bm{\mathcal{P}}_v(\phi) + \bm{K}(E) \big]^{-1}
\bm{M}^{-1/2},
\end{equation}
where $\bm{K}(E)$ is the diagonal matrix in eq.~\eqref{eq:Kmatrix}. The limit must be taken after evaluating the inverse of the matrix between the factors of $\bm{M}^{-1/2}$ in eq.~\eqref{eq:AsmatrixLOC}. The matrix $ \bm{\mathcal{A}}(E)$  in eq.~\eqref{eq:A-sC} reduces to
\begin{equation}
\label{eq:AmatrixLOC}
\bm{\mathcal{A}}(E) = 
\begin{pmatrix} ~0~  & 0  \\  0  & \mathcal{A}_C(E) \end{pmatrix} 
+\frac{8\pi}{L_u(E)}  \, 
\begin{pmatrix} ~1~  & 0  \\  0  & W_1(E) \end{pmatrix} 
\bm{M}^{-1/2} \, \bm{\mathcal{P}}_u(\phi)\,  \bm{M}^{-1/2}
\begin{pmatrix} ~1~  & 0  \\  0  & W_1(E) \end{pmatrix} .
\end{equation}
The denominator in the second term is
\begin{equation}
L_u(E)=-\gamma_u +\cos^2\phi\,  \kappa_0(E)  +\sin^2\phi \, K_1(E),
\label{eq:Ku}
\end{equation}
where $\kappa_0(E)$ is given in eq.~\eqref{eq:kappa0} and $K_1(E)$ is given in eq.~\eqref{eq:K1-E}.

The neutral-wino scattering length $a_0$ can be obtained by evaluating the transition amplitude $\mathcal{A}_{00}(E)$ at the neutral-wino-pair threshold:
\begin{equation}
\mathcal{A}_{00}(E=0) = - 8\pi a_0 /M .
\label{eq:T00-a0}
\end{equation}
The inverse neutral-wino scattering length $\gamma_0 \equiv 1/a_0$ is
\begin{equation}
\gamma_0 = (1 +  t_\phi^2)\gamma_u - t_\phi^2\, K_1(0) ,
\label{eq:a0-gammauLO}
\end{equation}
where $t_\phi \equiv \tan \phi$. This equation can be solved for $\gamma_u$ as a function of $\gamma_0$:
\begin{equation}
\gamma_u = \frac{t_\phi^2\, K_1(0)  + \gamma_0}{1 +  t_\phi^2}.
\label{eq:gammau-a0LO}
\end{equation}
If $\gamma_0 = 0$, the elastic neutral-wino cross section $\sigma_{0 \to 0}(E)$ saturates the unitarity bound in eq.~\eqref{sigma-unitarity0} in the limit $E \to 0$. For this reason, we refer to the critical value $\gamma_0 = 0$ as {\it unitarity}. If $|\gamma_0| \ll \sqrt{2M\delta}$, there are large cancellations in the denominator $L_u(E)$ in eq.~\eqref{eq:Ku}. These cancellations can be avoided by eliminating $\gamma_u$ in favor of $\gamma_0$. The resulting expression for the matrix of transition amplitudes is 
\begin{equation}
\label{eq:AmatrixLOC1}
\bm{\mathcal{A}}(E) = 
 \begin{pmatrix} ~0~ & 0 \\ 0 & {\cal A}_C(E)
\end{pmatrix} 
 + \frac{8\pi}{L_0(E)} \begin{pmatrix} 1  & 0 \\  0 & W_1(E) \end{pmatrix} \bm{M}^{-1/2}   
\begin{pmatrix}    ~1~     & t_\phi\\  t_\phi & t_\phi^2 \end{pmatrix} 
\bm{M}^{-1/2} \begin{pmatrix} ~1~   & 0 \\  0 & W_1(E) \end{pmatrix},
\end{equation}
where ${\cal A}_C(E)$ is the Coulomb amplitude in eq.~\eqref{eq:ACoulomb} and $W_1(E)$ is the amplitude in eq.~\eqref{eq:S-eta} for $w^+ w^-$ created at a point to become  $w^+ w^-$ with energy $E$. The denominator in the second term is
\begin{equation}
L_0(E) =-\gamma_0  + t_\phi^2\, \big[ K_1(E) - K_1(0) \big]\, + \kappa_0(E),
\label{eq:L0-E}
\end{equation}
where $\kappa_0(E)$ is given in eq.~\eqref{eq:kappa0} and the function $K_1(E)$ is given in eq.~\eqref{eq:K1-E}.

\subsection{Wino-wino scattering}
\label{sec:CrossSectionLO}

The cross section for elastic scattering from channel $i$ to channel $j$ at energy $E$, averaged over initial spins and summed over final spins, is denoted by  $\sigma_{i \to j}(E)$. The expressions for these cross sections in terms of the T-matrix elements ${\cal T}_{ij}(E)$ for states with the standard normalizations of a nonrelativistic field theory are
\begin{subequations}
\begin{eqnarray}
\sigma_{i \to 0}(E) &=&\frac{M^2}{8\pi}
\big| {\cal T}_{i0}(E) \big|^2 \frac{v_0(E)}{v_i(E)},
\label{eq:sig0E-calT}
\\
\sigma_{i\to 1}(E) &=&\frac{M^2}{4\pi}
\big| {\cal T}_{i1}(E) \big|^2 \frac{v_1(E)}{v_i(E)},
\label{eq:sig1E-calT}
\end{eqnarray}
\label{eq:sigE-calT}%
\end{subequations}
where $v_i(E)$ and $v_j(E)$ are the velocities of the incoming and outgoing winos, which are given in eqs.~\eqref{eq:v0,1-E}. The extra factor of $1/2$ in the cross sections  $\sigma_{i \to 0}$ in eq.~\eqref{eq:sig0E-calT} for producing a neutral-wino pair compensates for overcounting by integrating over the entire phase space of the two identical particles. The T-matrix elements ${\cal T}_{ij}(E)$ are obtained by evaluating the transition amplitudes $\mathcal{A}_{ij}(E)$ on the appropriate energy shell. For a neutral-wino pair $w^0 w^0$ with relative momentum $p$, the energy shell is $E = p^2/M$. For a charged-wino pair $w^+ w^-$ with relative momentum $p$, the energy shell is $E = 2 \delta + p^2/M$.

For center-of-mass energy in the range $0 \leq E < 2\delta$ below the charged-wino-pair threshold, only the neutral-wino-pair channel is open. The T-matrix element for  $w^0w^0 \to w^0w^0$ in ZREFT at LO is given by the 00 entry of the matrix in eq.~\eqref{eq:AmatrixLOC1}:
\begin{equation}
{\cal T}_{00}(E) = \frac{8\pi/M}{L_0(E)},
\label{eq:T00LO}
\end{equation}
where $L_0(E)$ is given in eq.~\eqref{eq:L0-E}. The reciprocal of the T-matrix element ${\cal T}_{00}(E)$ for neutral-wino elastic scattering can be expanded in powers of the relative momentum $p = \sqrt{ME}$:
\begin{equation}
\frac{8\pi/M}{{\cal T}_{00}(E)} =
-\gamma_0  - i p + \tfrac12 r_0 \, p^2+ \tfrac18 s_0 \, p^4  + {\cal O}(p^6).
\label{eq:T00NLOinv}
\end{equation}
The only odd power of $p$ in the expansion is the pure imaginary term $-ip$. The coefficients of the even powers of $p$ are real valued. The leading term $-\gamma_0$ vanishes at unitarity. The effective range  $r_0$ and the shape parameter $s_0$ can be determined by expanding the real part of  $1/{\cal T}_{00}(E)$ from eq.~\eqref{eq:T00LO} in powers of $p^2$ and comparing to the expansion in eq.~\eqref{eq:T00NLOinv}:
\begin{subequations}
\begin{eqnarray}
r_0 &=& 2 t_\phi^2\, K_1'(0)/M,
\label{eq:r0LO}
\\
s_0 &=& 4 t_\phi^2\, K_1''(0)/M^2.
\label{eq:s0LO}
\end{eqnarray}
\label{eq:r0s0LO}%
\end{subequations}
The predictions for these coefficients are  independent of $\gamma_0$.

For energy in the range $ E > 2\delta$ above the charged-wino-pair threshold, the $w^0\, w^0$ and $w^+ w^-$ channels are both open. The T-matrix elements  in ZREFT at LO for  $w^0w^0 \to w^0w^0$ is given in eq.~\eqref{eq:T00LO}. The T-matrix elements  in ZREFT at LO for  $w^0w^0 \to w^+w^-$ and $w^+w^- \to w^+w^-$ are given by the 01 and 11 entries of the matrix in eq.~\eqref{eq:AmatrixLOC1}:
\begin{subequations}
\begin{eqnarray}
{\cal T}_{01}(E) &=& \frac{(4\sqrt2\, \pi/M) t_\phi W_1(E)}{L_0(E)},
\label{eq:T01LO}
\\
{\cal T}_{11}(E) &=& \mathcal{A}_C(E) + \frac{(4 \pi/M) t_\phi^2 W_1^2(E)}{L_0(E)},
\label{eq:T11LO}
\end{eqnarray}
\label{eq:T01,11LO}%
\end{subequations}
where $L_0(E)$ is given in eq.~\eqref{eq:L0-E}, $W_1(E)$ is given in eq.~\eqref{eq:S-eta}, and $\mathcal{A}_C(E)$ is the on-shell Coulomb amplitude in eq.~\eqref{eq:ACoulomb}.

\subsection{Matching with NREFT}
\label{sec:MatchingLO}

The interaction parameters of ZREFT can be determined by matching T-matrix elements in ZREFT with low-energy T-matrix elements in NREFT. The dimensionless T-matrix elements $T_{ij}(E)$ for wino-wino scattering in NREFT can be calculated numerically by solving the coupled-channel Schr\"odinger equation in eq.~\eqref{eq:radialSchrEq}. The T-matrix elements ${\cal T}_{ij}(E)$ for wino-wino scattering in ZREFT at LO are given analytically  in eqs.~\eqref{eq:T00LO} and \eqref{eq:T01,11LO}. For $E>2 \delta$, the relation between the T-matrix in NREFT and the T-matrix in ZREFT is \cite{Braaten:2017gpq}
\begin{equation}
\frac{1}{2M} \, \bm{v}(E)^{-1/2} \, \bm{T}(E)  \, \bm{v}(E)^{-1/2}= 
\frac{1}{8\pi} \, \bm{M}^{1/2} \, \bm{\mathcal{T}}(E) \, \bm{M}^{1/2},
\label{eq:TNR-TZR}
\end{equation}
where $\bm{M}$ is the diagonal matrix of masses in eq.~\eqref{eq:Mmatrix} and $\bm{v}(E)$ is the diagonal matrix of the velocities defined in eq.~\eqref{eq:v0,1-E}:
\begin{equation}
\bm{v}(E) = 
\begin{pmatrix} v_0(E) & 0 \\ 0 & v_1(E) \end{pmatrix}.
\label{eq:vmatrix}
\end{equation}
For $0<E<2 \delta$, the relation between the T-matrix elements for neutral-wino scattering is
\begin{equation}
\frac{1}{2Mv_0(E)} T_{00}(E) = \frac{M}{8 \pi}\, {\cal T}_{00}(E).
\label{eq:TNR-TZR00}
\end{equation}

The interaction parameters of ZREFT at LO are $\alpha$, $\phi$, and $\gamma_0=1/a_0$. An accurate parametrization of neutral-wino scattering length $a_0(M)$ for NREFT with $\delta = 170$~MeV and $M$ near the critical mass $M_*$ is provided by the Pad\'e approximant in eq.~\eqref{eq:a0Pade}. The angle $\phi$ can be determined by matching some other physical quantity in ZREFT and in NREFT. If $\delta$ is fixed, it is better to use a value of $M$ close to the unitarity value  $M_*(\delta)$ and to match a T-matrix element at an energy $E$ close to 0. The expansion of the reciprocal of the T-matrix element ${\cal T}_{00}(E)$ for neutral-wino elastic scattering in powers of the relative momentum $p = \sqrt{ME}$ is given in eq.~\eqref{eq:T00NLOinv}. The corresponding expansion in powers of $p$  in NREFT is given in eq.~\eqref{eq:T00NRinv}. At unitarity, the lowest-energy quantity that can be used for matching is the effective range $r_0$.

If we choose the effective range at some mass $M$ as the matching quantity, the matching condition for ZREFT at LO is
\begin{equation}
t^2_\phi(M) = -  \frac{\Delta/2}{z_0^2\, \psi'(z_0) - \frac12 - z_0}r_0(M),
\label{eq:r0match}
\end{equation}
where $\Delta = \sqrt{2 M \delta}$ and $z_0 = - \alpha M/(2 \Delta)$. In the limit $\alpha \to 0$, the matching condition reduces to $r_0 = - \tan^2\phi/\Delta$. The derivative of $\psi(z)$ can be expanded as a power series that converges for $|z| < 1$:
\begin{equation}
\psi'(z) = \frac{1}{z^2} + \sum_{n=0}^\infty (-1)^n (n+1) \,\zeta(n+2)\, z^n,
\label{eq:psi'-z}
\end{equation}
where $\zeta(z)$ is the Riemann  zeta function. This can be used to expand the right side of eq.~\eqref{eq:r0match} as a power series in  $z_0 = - \alpha M/(2 \Delta)$. The convergence rate of the expansion is determined not by the size of $\alpha=1/137$, but instead by the size of the ratio $\alpha M/(2 \Delta)$. For $\delta = 170$~MeV, the value of this ratio at unitarity is 0.306. Thus although the effective range provides a matching condition that is perturbative in $\alpha$, matching at $\alpha=0$ is not quantitatively useful.

\begin{figure}[t]
\centering
\includegraphics[width=0.8\linewidth]{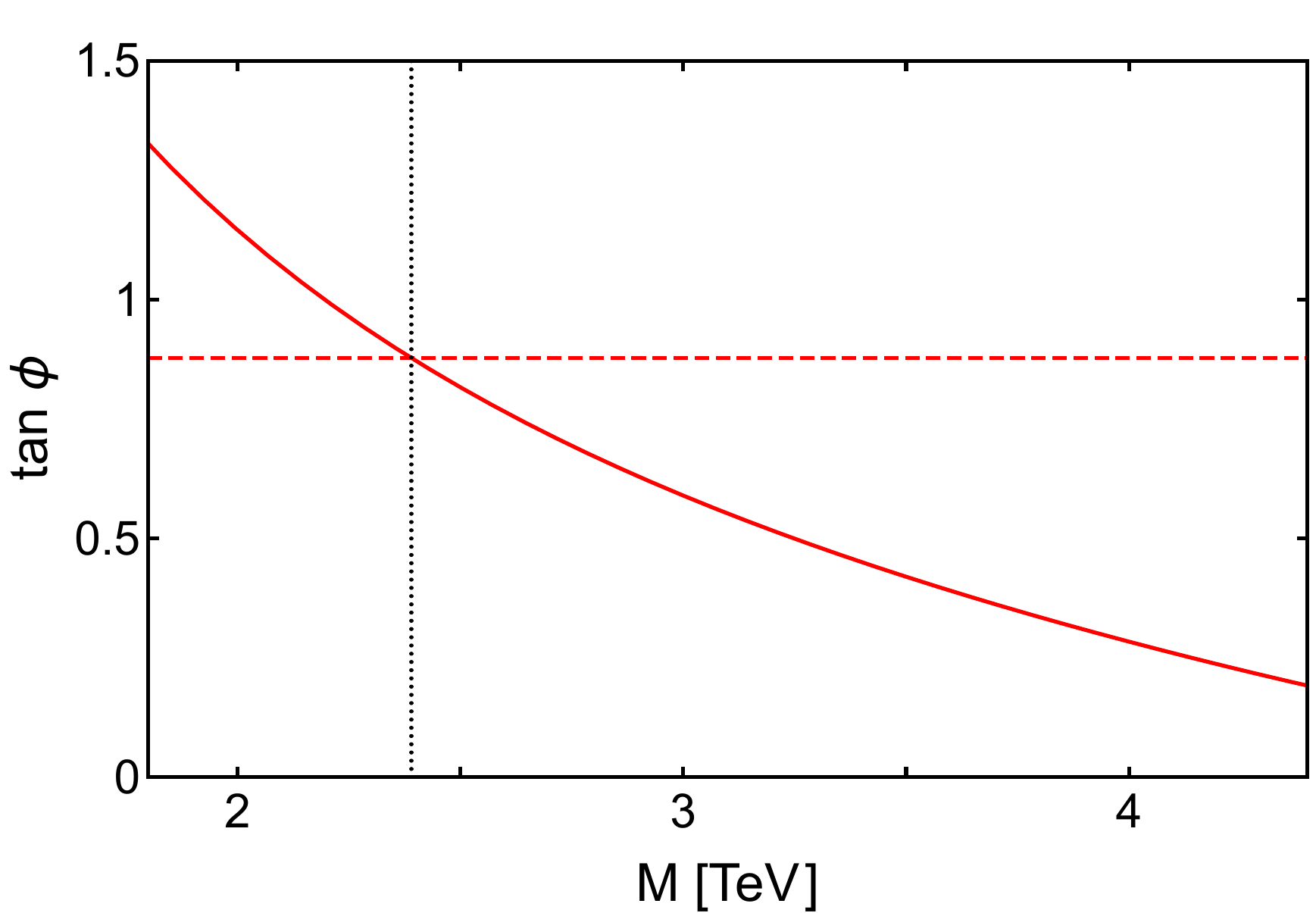}
\caption{Interaction parameter $\tan\phi$ for ZREFT at LO as a function of the wino mass $M$ for $\delta = 170$ MeV and $\alpha = 1/137$. The parameter $\tan\phi(M)$ (solid red curve) is determined from the matching condition for the effective range $r_0$ in NREFT in eq.~\eqref{eq:r0match}. The constant value $\tan\phi(M_*)=0.877$ (dashed red line) is determined by matching $r_0$ at unitarity. The vertical dotted line marks the unitarity mass $M_* = 2.39$~TeV.}
\label{fig:tanphivsM}
\end{figure}

A specific choice for the matching point in eq.~\eqref{eq:r0match} is $\alpha = 1/137$, $\delta =170$~MeV, and the unitarity mass $M_* = 2.39$~TeV. Using the numerical result for the effective range in NREFT  in eq.~\eqref{eq:r0*EM}, our matching condition in eq.~\eqref{eq:r0match} gives $\tan\phi = 0.877$, where we have chosen the positive root. The mixing angle $\phi$ is about $40^\circ$. A different choice for the matching mass $M$ near $M_*$ would give a  different value for $\tan \phi$. For $M$ near $M_*$, the effective range $r_0(M)$ can be accurately approximated by the Pad\'e approximant in eq.~\eqref{eq:r0Pade}. The value of $\tan \phi(M)$ determined by inserting this Pad\'e approximant into the matching condition in eq.~\eqref{eq:r0match} is shown as a function of  $M$ in figure~\ref{fig:tanphivsM}. It varies significantly with $M$ within the range of validity of ZREFT. In the predictions of ZREFT at LO at the mass $M$, it is therefore better to use the value of $\tan \phi(M)$ from matching at the mass $M$ than the value  $\tan\phi(M_*) = 0.877$ from matching at unitarity.

\begin{table}[t]
\begin{center}
\begin{tabular}{|ccc|cc|c|}
\hline
$\alpha$ & ~$M$~(TeV)~ & ~$\delta$~(MeV)~ & ~$\gamma_0 /(2M\delta)^{1/2}$~ & ~$r_0 (2M\delta)^{1/2}$~ & ~~$\tan\phi$~~ \\
\hline
~1/137~  &   2.39~~~   & 170 &     0     &   ~$-1.653$~   &  0.877   \\
        0    &   2.39~~~   & 170 &  $-1.277$  &    $-1.224$      & 1.106   \\ 
        0    &   2.88~~~   & 170 &     0     &     $-0.693$     & 0.832   \\
        0    &   2.22~~~   &   0  &     0     &     $-1.552$     & 1.246   \\
\hline
\end{tabular}
\end{center}
\caption{Interaction parameter $\tan\phi$ for ZREFT at LO from matching to NREFT at various matching points. The inverse scattering length $\gamma_0$ and the effective range $r_0$ are calculated using NREFT. The parameter $\tan\phi$  is determined by the matching condition for $r_0$ in eq.~\eqref{eq:r0match}.}
\label{tab:Parameters}
\end{table}

If $\alpha$ was small enough, we could determine $\tan \phi$ by matching predictions from ZREFT with $\alpha = 0$ to results from NREFT with $\alpha = 0$. The effective range for $\alpha = 0$, $\delta = 170$~MeV, and $M_*=2.39$~TeV is given in the text after eq.~\eqref{eq:r0,s0*EM}. By matching it to the prediction $r_0 = -\tan^2\phi/\Delta$ from ZREFT at LO with $\alpha = 0$, we obtain $\tan \phi =1.106$. This value of $\tan \phi$ is listed in table~\ref{tab:Parameters}, along with the values obtained in ref.~\cite{Braaten:2017gpq} at two other matching points with $\alpha=0$. Significant differences in the value of $\tan \phi$ imply significant differences in the predictions of ZREFT at LO. Matching at  $\alpha =0$, $\delta = 170$~MeV, and the corresponding unitarity mass $M_*=2.88$~TeV gives a value of $\tan \phi$ that is only 5\% lower than that from matching at $\alpha = 1/137$, $\delta = 170$~MeV, and the unitarity mass $M_*=2.39$~TeV.

\subsection{Predictions of ZREFT at LO}
\label{sec:PredictLO}

\begin{figure}[t]
\centering
\includegraphics[width=0.48\linewidth]{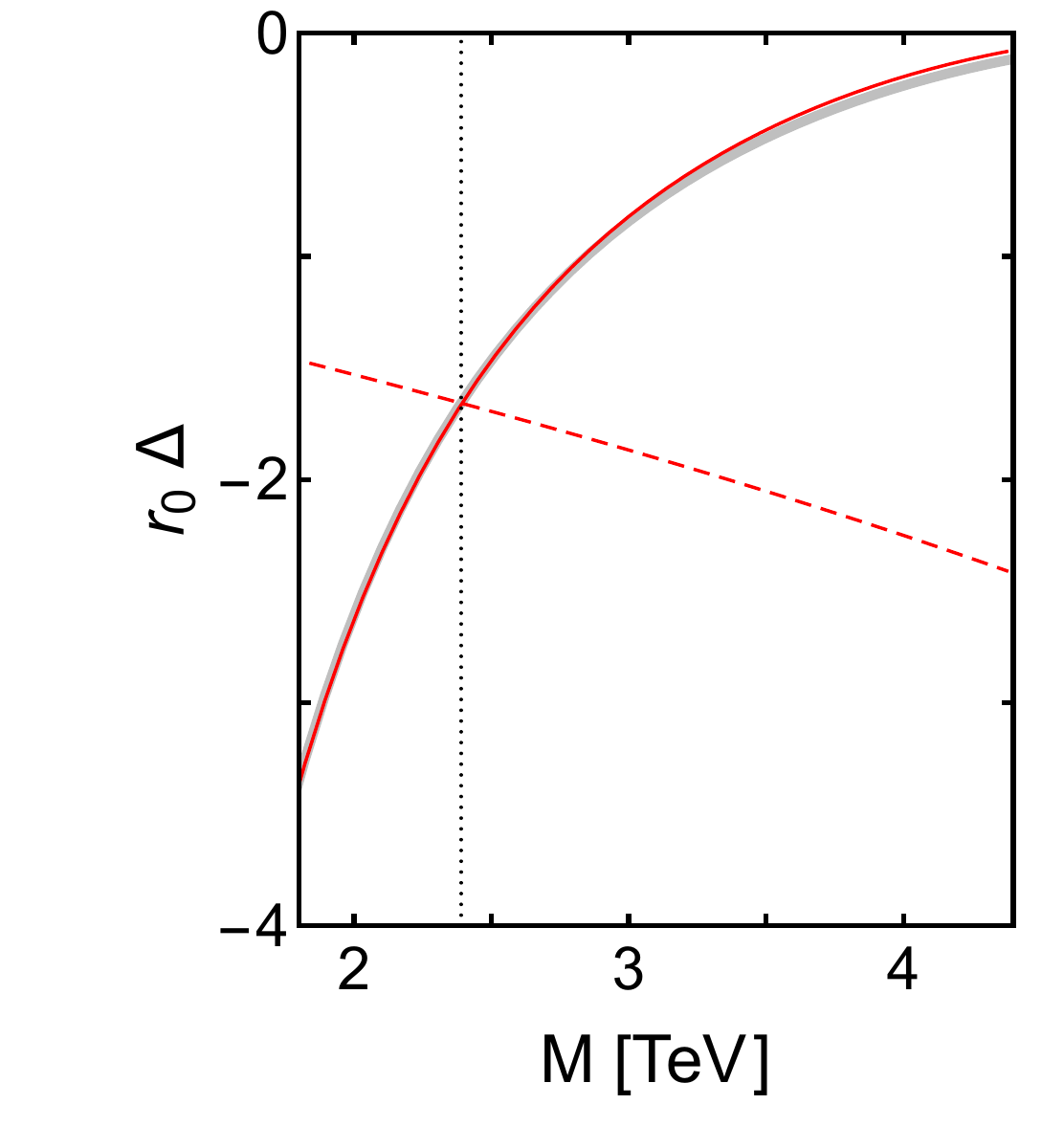}
\includegraphics[width=0.48\linewidth]{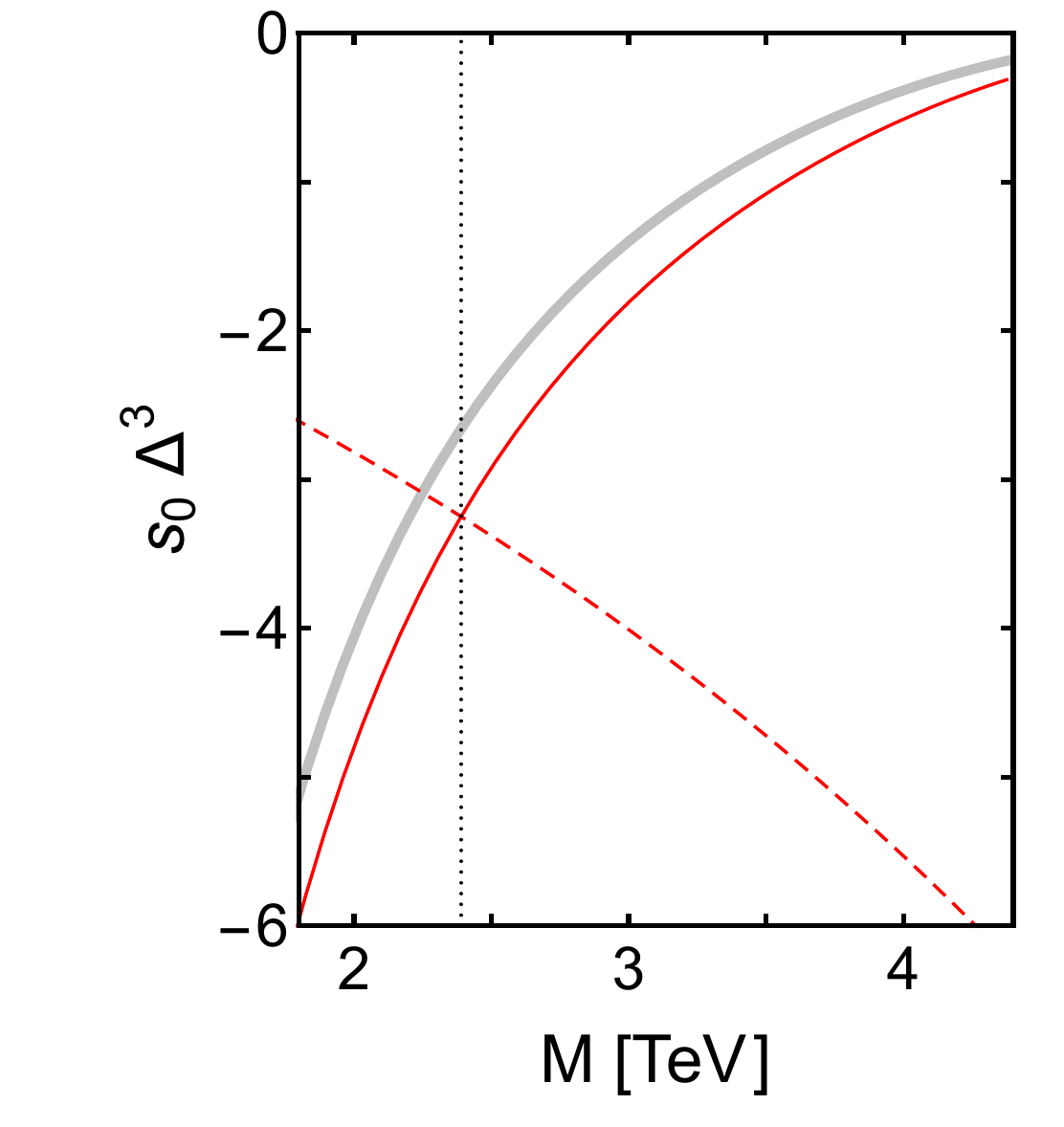}
\caption{Neutral-wino effective range $r_0$ (left panel) and shape parameter $s_0$ (right panel) as functions of the wino mass $M$: NREFT (thicker grey curve), ZREFT at LO  with $\tan \phi(M)$ (solid red curve), and ZREFT at LO with  $\tan \phi(M_*) = 0.877$ (dashed red curve). The vertical dotted lines indicate the  unitarity mass $M_* = 2.39$~TeV.}
\label{fig:r0s0vsM}
\end{figure}

The predictions of ZREFT at LO as a function of the wino mass $M$ can be obtained by using the Pad\'e approximant for the inverse scattering length $\gamma_0(M)= 1/a_0(M)$ given by eq.~\eqref{eq:a0Pade} and the $M$-dependent mixing angle $\phi(M)$ determined by inserting the Pad\'e approximant for the effective range $r_0(M)$ in eq.~\eqref{eq:r0Pade} into the matching condition in eq.~\eqref{eq:r0match}. As shown in the left panel of figure~\ref{fig:r0s0vsM}, the Pad\'e approximant for $r_0(M)$ is very accurate over the entire range of validity of ZREFT. In the right panel of figure~\ref{fig:r0s0vsM}, the prediction of ZREFT at LO  for the shape parameter $s_0(M)$ as a function of $M$ is compared to the result from NREFT. At unitarity, the prediction for  $s_0$ differs from the result from NREFT in eq.~\eqref{eq:s0*EM} by a multiplicative factor of 0.81. The accuracy of the prediction remains comparable at other values of $M$ within the range of  validity of ZREFT. If $M$ is very close to the unitarity mass $M_* = 2.39$~TeV, we can use the constant mixing angle given by $\tan \phi(M_*)=0.877$, which was determined by matching $r_0$ at unitarity. However, as shown in figure~\ref{fig:r0s0vsM}, the resulting predictions  for $r_0(M)$ and $s_0(M)$ as functions of $M$ have the wrong slopes. The accuracy of the predictions therefore deteriorates quickly as $|M-M_*|$ increases.

The energy dependence of the wino-wino cross sections is most dramatic at a unitarity mass. ZREFT at LO can be applied at the unitarity mass $M_* = 2.39$~TeV by setting $\gamma_0 = 0$ and by setting $\tan \phi = 0.877$.  We compare the energy dependence of the wino-wino cross sections predicted by ZREFT at LO with Coulomb resummation with the results from NREFT.

\begin{figure}[t]
\centering
\includegraphics[width=0.8\linewidth]{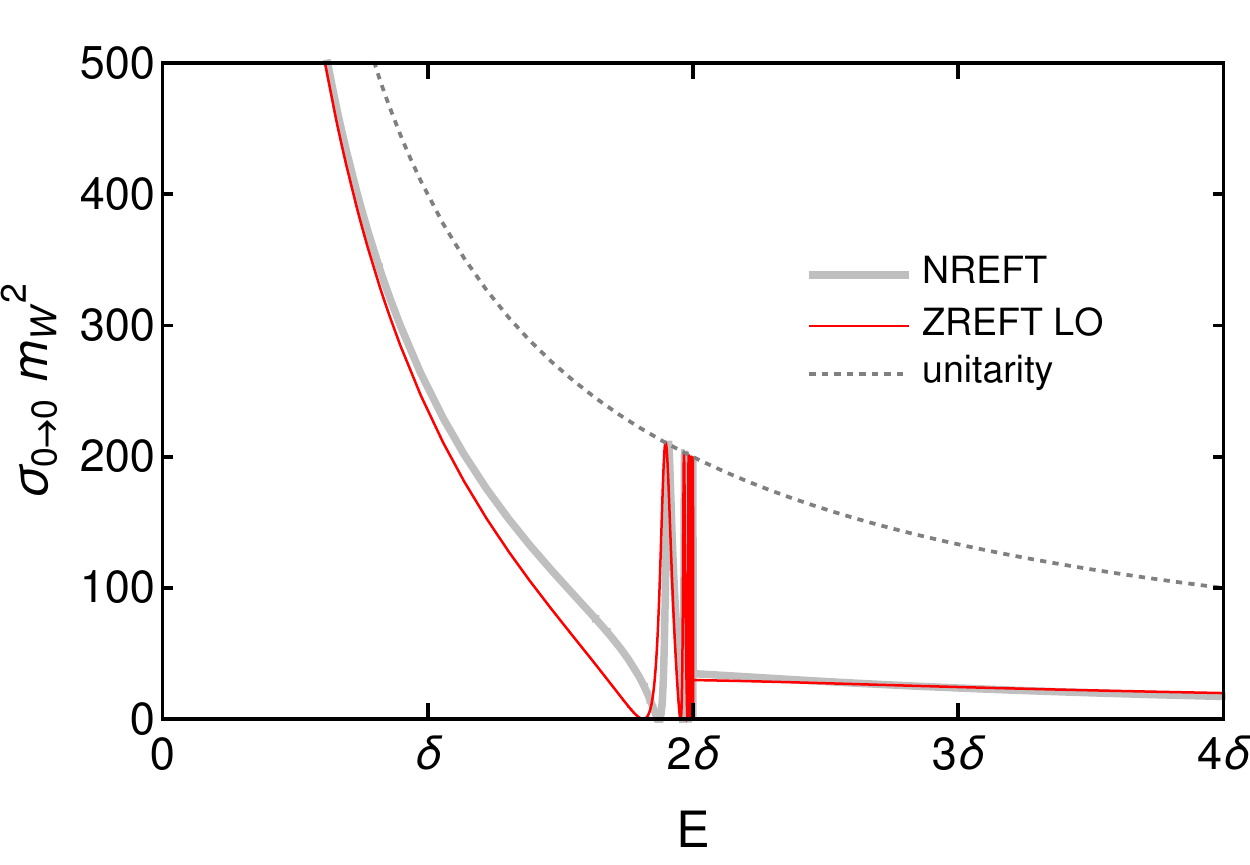}
\caption{Neutral-wino elastic cross section $\sigma_{0 \to 0}$ as a function of the energy $E$. The cross section at the unitarity mass $M_*=2.39$~TeV is shown for NREFT (thicker grey curve) and for ZREFT at LO with $\tan \phi = 0.877$ (red curve). The S-wave unitarity bound is shown as a dotted curve.}
\label{fig:sigma00vsE-LO}
\end{figure}

\begin{figure}[t]
\centering
\includegraphics[width=0.49\linewidth]{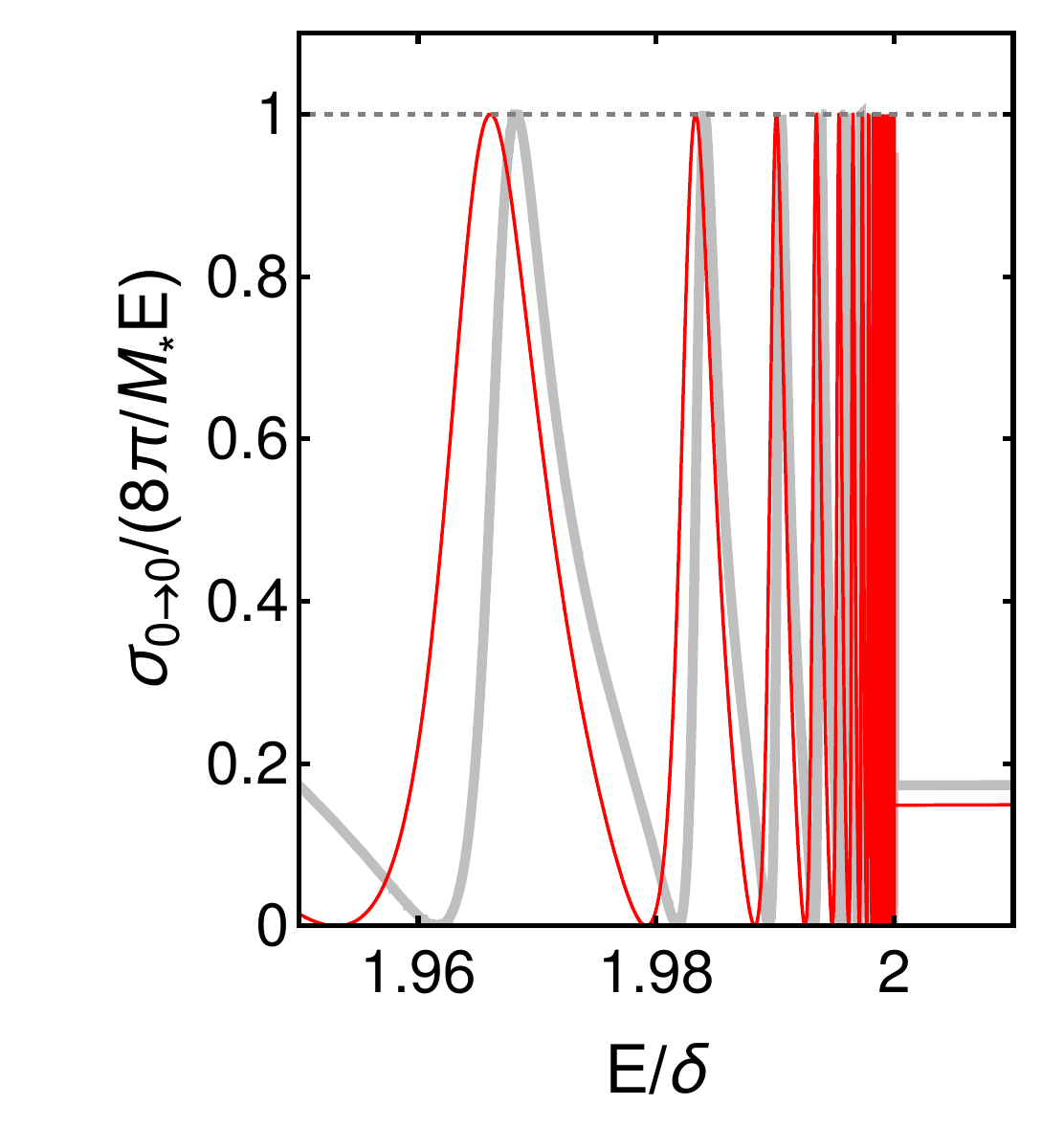}
\caption{Neutral-wino elastic cross section $\sigma_{0 \to 0}$ divided by the S-wave unitarity bound as a function of the energy $E$ near the charged-wino-pair threshold. The cross section at the unitarity mass $M_*=2.39$~TeV is shown for NREFT (thicker grey curve) and for ZREFT at LO with $\tan \phi = 0.877$ (red curve).}
\label{fig:sigma00vsE-LO_th}
\end{figure}

In figure~\ref{fig:sigma00vsE-LO}, we compare the energy dependence of the cross section for neutral-wino elastic scattering predicted by ZREFT at LO  with $\tan \phi = 0.877$  with the results from NREFT. In the limit $E \to 0$, both cross sections saturate the unitarity bound. The mixing angle $\phi(M_*)$  was tuned so that the next-to-leading term in the low-energy expansions also agrees. The prediction of ZREFT at LO also agrees well with the result from NREFT above the charged-wino-pair threshold at $2 \delta$. Just above $2\delta$, the prediction is smaller by a factor of 0.857. The prediction for $\sigma_{0 \to 0}$ at $E > 2 \delta$ can be improved by decreasing $\tan\phi$ at the cost of decreasing the accuracy of the prediction for $E$ close to 0. The mixing angle determined by matching $\sigma_{0 \to 0}$ in the limit $E \to 2 \delta^+$ is given by $\tan \phi = 0.825$. There are significant differences between the prediction of ZREFT at LO and the results from NREFT in the resonance region just below the threshold at $2 \delta$. A blow-up of the threshold region, with the cross section divided by the S-wave unitarity bound, is shown in figure~\ref{fig:sigma00vsE-LO_th}. Just below the threshold, there is a sequence of increasingly narrow resonances associated with Coulomb $w^+ w^-$ bound states that saturate the unitarity bound. ZREFT at LO reproduces the qualitative behavior of the dramatic energy dependence. It predicts that the cross section has zeros and resonant peaks at the energies where the real part of the function $L_0(E)$ in eq.~\eqref{eq:L0-E} has poles and zeros, respectively. At unitarity where $\gamma_0 = 0$, the predictions for the zeros and resonant peaks are independent of the mixing angle $\phi$. The zeros in the cross section are predicted to be at the energies $E_n$ of the Coulomb bound states in eq.~\eqref{eq:E-n}. More accurate predictions for $\sigma_{0 \to 0}$ in the resonance region could be obtained by using ZREFT at NLO, which has two additional relevant parameters.

\begin{figure}[t]
\centering
\includegraphics[width=0.48\linewidth]{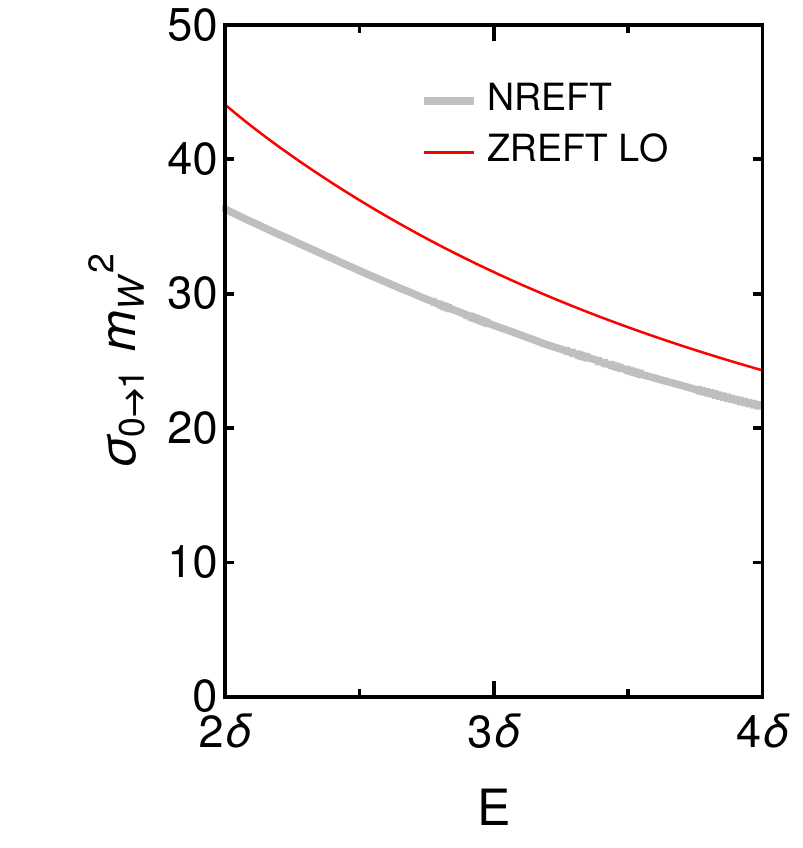}
\includegraphics[width=0.48\linewidth]{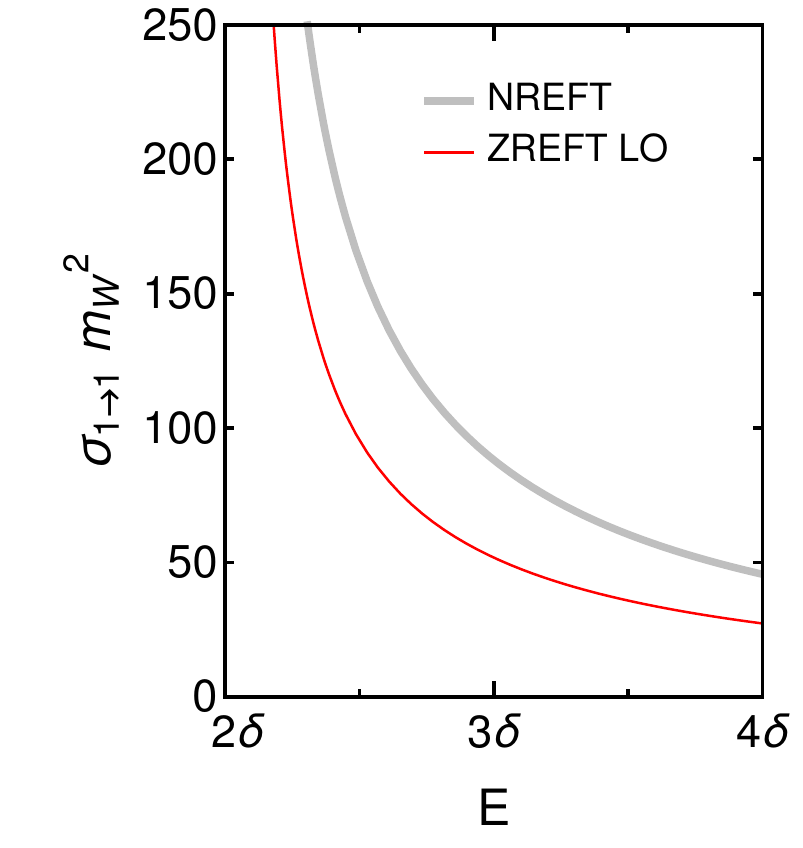}
\caption{Neutral-to-charged transition cross section $\sigma_{0 \to 1}$  (left panel) and the charged-wino elastic cross section $\sigma_{1 \to 1}$  (right panel) as functions of the energy $E$. The cross sections at the unitarity mass $M_*=2.39$~TeV are shown for NREFT (thicker grey curve)  and for ZREFT at LO  with $\tan \phi = 0.877$ (red curve).}
\label{fig:sigma01,11vsE-LO}
\end{figure}

In the left panel of figure~\ref{fig:sigma01,11vsE-LO}, we compare the energy dependence of the cross section for the neutral-to-charged transition predicted by ZREFT at LO with $\tan \phi = 0.877$ with the results from NREFT. The prediction has the correct qualitative behavior. It is larger by a factor that decreases from about 1.2 at the threshold to about 1.1 at $E=4 \delta$. The prediction for $\sigma_{0 \to 1}(E)$ can be improved by increasing  $\tan\phi$ at the cost of decreasing the accuracy of the prediction for $\sigma_{0 \to 0}$. The mixing angle determined by matching the cross section $\sigma_{0 \to 1}$ at $E=2\delta$ is given by $\tan \phi = 1.134$. The prediction at $E=4 \delta$ then differs from the result of NREFT by a factor of 0.99.

\begin{figure}[t]
\centering
\includegraphics[width=0.49\linewidth]{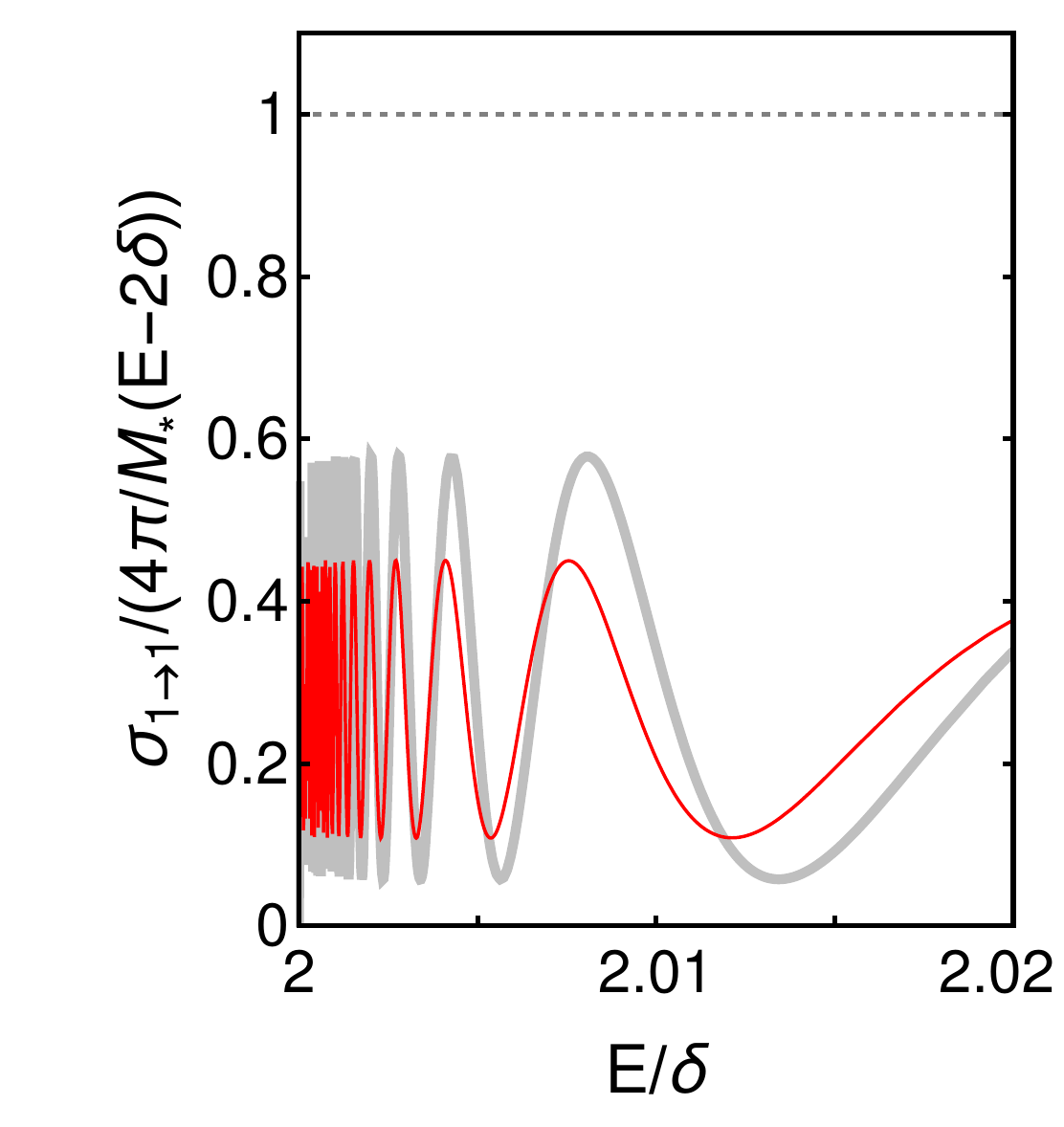}
\caption{Charged-wino elastic cross section $\sigma_{1 \to 1}$ divided by the S-wave unitarity bound as a function of the energy $E$ near the charged-wino-pair threshold. The cross section at the unitarity mass $M_*=2.39$~TeV is shown for NREFT (thicker grey curve) and for ZREFT at LO with $\tan \phi = 0.877$ (red curve).}
\label{fig:sigma11vsE-LO_th}
\end{figure}

In the right panel of figure~\ref{fig:sigma01,11vsE-LO}, we compare the cross section for charged-wino elastic scattering predicted by ZREFT at LO  with $\tan \phi = 0.877$ with the results from NREFT. The prediction seems to have the correct qualitative behavior. It is smaller than the NREFT cross section, differing by a factor that decreases to about 0.6 at $E=4 \delta$. The prediction for $\sigma_{1 \to 1}$ can be improved by increasing $\tan\phi$ at the cost of decreasing the accuracy of the prediction for $\sigma_{0 \to 0}$. Very near the charged-wino-pair threshold, both cross sections have dramatic oscillations that are too large to be visible in the right panel of figure~\ref{fig:sigma01,11vsE-LO}. A blow-up of the threshold region, with the cross sections divided by the S-wave unitarity bound, is shown in figure~\ref{fig:sigma11vsE-LO_th}. As $E$ approaches $2\delta$, the oscillations become increasingly narrow. They are not resonances, because they do not saturate the unitarity bound. The prediction of ZREFT at LO has the correct qualitative behavior. The predicted oscillations have amplitude smaller by about a factor of 0.66 and average value smaller by about a factor of 0.87. More accurate predictions for $\sigma_{1 \to 1}$  in the oscillation region could be obtained by using ZREFT at NLO, which has two additional relevant parameters.

\subsection{Wino-pair bound state}
\label{sec:BoundStateLO}

If the wino mass $M$ is larger than the unitarity mass where the neutral-wino scattering length $a_0(M)$ diverges, the S-wave resonance is a bound state below the neutral-wino-pair threshold. The bound state is a superposition of a neutral-wino pair and a charged-wino pair, and we denote it by $(ww)$. The coupled-channel radial Schr\"odinger equation for NREFT  in eq.~\eqref{eq:radialSchrEq} has a negative eigenvalue $- E_{(ww)}$, where $E_{(ww)}$ is the binding energy. In figure~\ref{fig:bindingenergyLO}, the binding energy for $\delta=170$~MeV is shown as a function of $M$. The binding energy goes to zero as $M$ approaches the unitarity mass $M_*=2.39$~TeV from above. The binding energy depends sensitively on the electromagnetic coupling constant $\alpha$. If the Coulomb potential between the charged winos is turned off by setting $\alpha=0$, the unitarity mass where $E_{(ww)}$ vanishes is shifted to 2.88~TeV.

\begin{figure}[t]
\centering
\includegraphics[width=0.8\linewidth]{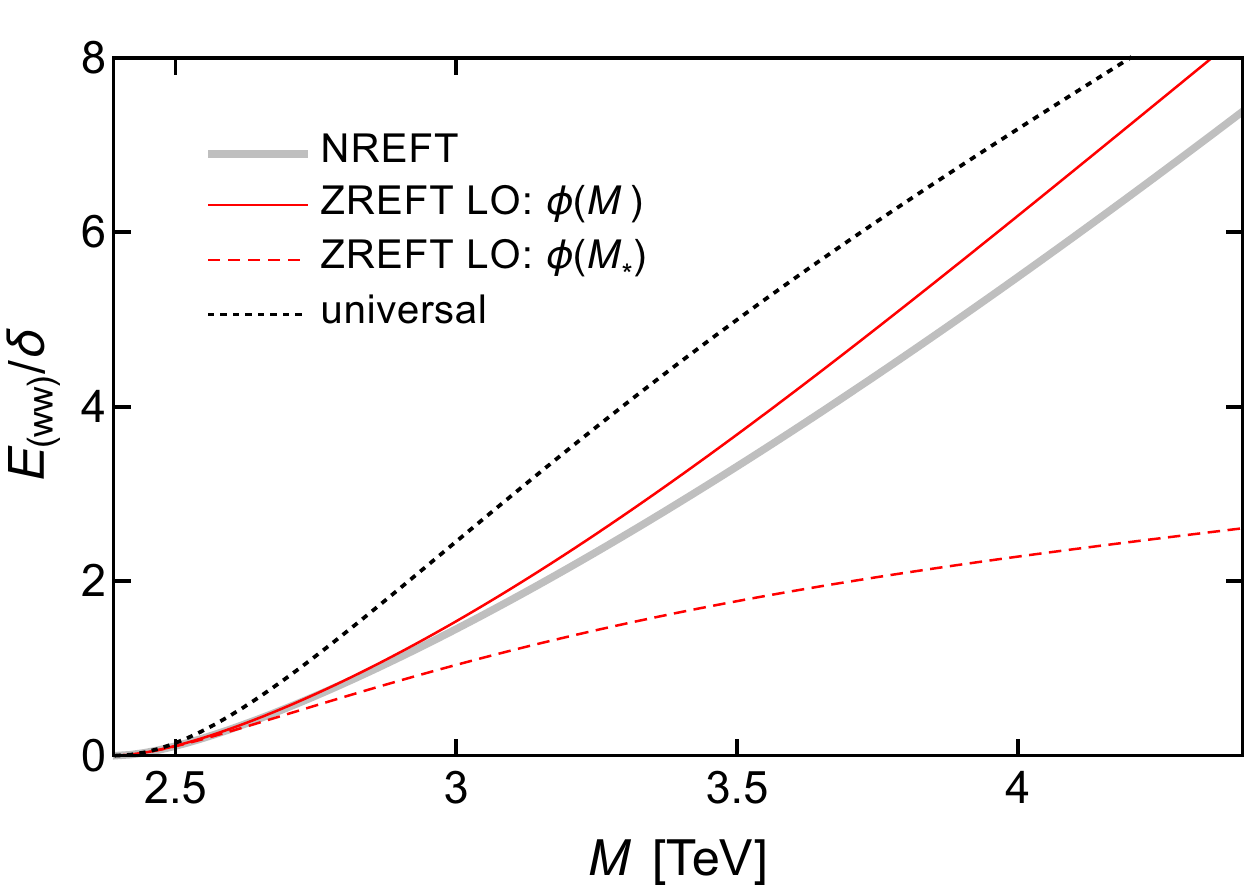}
\caption{Binding energy $E_{(ww)}$ of the wino-pair bound state as a function of the wino mass $M$: NREFT (thicker grey curve), ZREFT at LO with $\tan \phi(M)$ (red solid curve), ZREFT at LO with $\tan \phi(M_*) = 0.877$ (red dashed curve), and the universal approximation in eq.~\eqref{eq:Eww-largea} (dotted curve).}
\label{fig:bindingenergyLO}
\end{figure}

In ZREFT, the binding energy $E_{(ww)}$ of the wino pair bound state can be obtained by solving an analytic equation numerically. If $\gamma_0 > 0$, each of the transition amplitudes ${\cal A}_{ij}(E)$ given by the matrix in eq.~\eqref{eq:AmatrixLOC1} has a pole at a real energy  $-E_{(ww)}$ below the neutral-wino-pair threshold. The pole in $E$ is at a zero of the function $L_0(E)$ in eq.~\eqref{eq:L0-E}. The binding energy can be expressed as $E_{(ww)} = \gamma^2/M$, where the binding momentum $\gamma$ is a positive solution to the equation
\begin{equation}
0=  \gamma - \gamma_0   + t_\phi^2 \, \big[ K_1(-\gamma^2/M) -K_1(0) \big].
\label{eq:gammaLO-eq}
\end{equation}
The correct root of this equation is the one that approaches 0 as $\gamma_0$ decreases to 0 from above. In figure~\ref{fig:bindingenergyLO}, the predictions for the binding energy in ZREFT at LO are  compared to the result from NREFT. Using the $M$-dependent parameter $\tan \phi(M)$ obtained by matching $r_0(M)$ gives a prediction for $E_{(ww)}$ that tracks the result fairly well as a function of $M$. As $M \to M_*$, the prediction approaches the result from above. Its error decreases to less than 5\% for $M-M_* < 0.5$~TeV. Using the constant parameter $\tan \phi(M_*) = 0.877$ obtained by matching $r_0$ at unitarity gives a prediction for $E_{(ww)}$ whose error deteriorates quickly as $M$ increases. As $M \to M_*$, the prediction approaches the result from below. Its error decreases to less than 5\% for $M-M_* < 0.2$~TeV.

Particles with short-range interactions that produce an S-wave resonance sufficiently close to their scattering threshold have universal low-energy behavior that is completely determined by their S-wave scattering length $a_0$ \cite{Braaten:2004rn}. If $a_0$ is positive, the S-wave bound state closest to the threshold is universal. The universal approximation for its binding energy is
\begin{equation}
E_{(w w)} = 1/(M a_0^2).
\label{eq:Eww-largea}
\end{equation}
For neutral winos with mass near  the unitarity mass $M_*=2.39$~TeV, the universal approximation in eq.~\eqref{eq:Eww-largea} is applicable for $M$ inside the region between $M_*$ and 2.9~TeV. The universal approximation becomes increasingly accurate as $M$ approaches $M_*$. In figure~\ref{fig:bindingenergyLO}, the universal approximation in eq.~\eqref{eq:Eww-largea} with the Pad\'e approximant for $a_0(M)$ in eq.~\eqref{eq:a0Pade} is compared to the result from NREFT. As $M \to M_*$, the universal approximation approaches the result from above. Its error decreases to less than 5\% for $M-M_* < 0.004$~TeV.

The bound state $(ww)$ is a superposition of a neutral-wino  pair $w^0 w^0$ and a charged-wino pair $w^+ w^-$. The probabilities of the $w^0 w^0$ and $w^+ w^-$ components of the bound state $(ww)$  can be deduced from the transition amplitudes $\mathcal{A}_{00}(E)$ and $\mathcal{A}_{11}(E)$ in eq.~\eqref{eq:AmatrixLOC1}. Both of these amplitudes have a pole in the energy at  $E = - \gamma^2/M$, where $\gamma$ satisfies eq.~\eqref{eq:gammaLO-eq}. We denote the residues of the poles in $\mathcal{A}_{00}(E)$ and $\mathcal{A}_{11}(E)$ by $-\mathcal{Z}_0$ and $-\mathcal{Z}_1$, respectively. The absolute values of the residues $\mathcal{Z}_0$ and $\mathcal{Z}_1$ are proportional to the probabilities for the $w^0 w^0$ and $w^+ w^-$ components of the bound state, respectively. The residue factor for the $w^0 w^0$ channel at LO  is 
\begin{equation}
\mathcal{Z}_0 =
\frac{16 \pi \gamma/M^2}
       {1 - 2 t_\phi^2\,  K_1'(-\gamma^2/M) \, \gamma/M}.
\label{eq:Z0}
\end{equation}
The ratio of the residue factors at LO is
\begin{equation}
\mathcal{Z}_1/ \mathcal{Z}_0 = \frac12 t_\phi^2\,  W_1^2(-\gamma^2/M),
\label{eq:Z1/Z_0}
\end{equation}
where the function $W_1(E)$ is given in eq.~\eqref{eq:S-eta}. The ratio of the probabilities for $w^+ w^-$ and $w^0 w^0$ is 
\begin{equation}
\frac{|\mathcal{Z}_1|}{|\mathcal{Z}_0|/2} =
\tan^2\phi \, \Gamma^2(1-|\eta_0|),
\label{eq:|Z1/Z_0|}
\end{equation}
where $|\eta_0| = \alpha M/(2 \Delta)$. In the limit $\alpha \to 0$, the probability for $w^0 w^0$ reduces to $\cos^2\phi$. Given the numerical value $\tan\phi=0.877$ from the LO fit, the probability for $w^0 w^0$ is approximately 43\%.

\section{Summary}
\label{sec:Conclusion}

One of the options for a wimp is the neutral wino $w^0$, which belongs to an $SU(2)$ multiplet that also includes the charged winos $w^+$ and $w^-$. The splitting $\delta$ between a charged wino and a neutral wino is small compared to the mass $M$ of the wino. The physics of nonrelativistic winos involves many momentum scales, including 
\begin{itemize}
\item 
the weak gauge boson mass scale $m_W$,
\item 
the scale  $\alpha_2 M$  of nonperturbative effects from exchange of weak gauge bosons,
\item 
the Bohr momentum  $\alpha M$, which is the scale of nonperturbative effects from the Coulomb interaction,
\item 
the scale $\sqrt{2M\delta}$ associated with the transition between a neutral-wino pair and a charged-wino pair,
\item 
the inverse scattering length $\gamma_0 = 1/a_0$ of the neutral wino. 
\end{itemize}
A fundamental description of winos is provided by a relativistic quantum field theory. Nonrelativistic effective field theories provide simpler descriptions for low-energy winos in which some of the momentum scales are not described explicitly.

If the winos are nonrelativistic, the momentum scale $M$ does not need to be treated explicitly. The winos can be described by the nonrelativistic effective field theory called NREFT. In NREFT, low-energy winos interact instantaneously at a distance through a potential generated by the exchange of weak gauge bosons, and charged winos also have local couplings to the electromagnetic field. If $M$ is large enough that $\alpha_2 M$ is comparable to  $m_W$, interactions between nonrelativistic winos from the exchange of the $W^\pm$ and $Z^0$ are nonperturbative. The effects of Coulomb interactions  between charged winos are also nonperturbative. Calculations in NREFT then require the numerical solution of a coupled-channel Schr\"odinger equation. The power of NREFT has recently been demonstrated by a calculation of the capture rates of two neutral winos into wino-pair bound states through the radiation of a photon \cite{Asadi:2016ybp}.

There are critical values of the wino mass at which there is an S-wave resonance at the neutral-wino-pair threshold. We refer to such a critical value as a unitarity mass, because the cross section saturates the S-wave unitarity bound in the low-energy limit. If $M$ is near a unitarity mass, the inverse scattering length $\gamma_0$ is much smaller than the momentum scales $m_W$ and $\alpha_2 M$. If the relative momentum of winos is smaller than $m_W$ and $\alpha_2 M$, those momentum scales do not need to be described explicitly. In Ref.~\cite{Braaten:2017gpq}, we developed a zero-range effective field theory called ZREFT to describe winos with mass $M$ near a unitarity mass. The effects of the exchange of weak gauge bosons between winos is reproduced by zero-range interactions between the winos that must be treated nonperturbatively. Charged winos also have local couplings to the electromagnetic field. The effects of Coulomb interactions between charged winos must also be  treated nonperturbatively. The power of ZREFT was illustrated in Ref.~\cite{Braaten:2017gpq} by calculating the rate for the formation of the wino-pair bound state in the collision of two neutral winos through a double radiative transition in which two soft photons are emitted.

NREFT is more broadly applicable than ZREFT. NREFT can describe nonrelativistic winos with any mass $M$, while ZREFT is only applicable if the wino mass is in a window around a unitarity mass. If the wino mass splitting is $\delta = 170$~MeV, the first such unitarity mass is $M_* = 2.39$~TeV, and the window for the applicability of ZREFT is $M$ from about 1.8~TeV to about 4.6~TeV. NREFT describes nonrelativistic winos, while ZREFT can only describe winos with relative momentum less than $m_W$. NREFT can describe the interactions of a pair of winos in any angular-momentum channel, while ZREFT can only describe S-wave interactions. Despite its more limited applicability, ZREFT has distinct advantages over NREFT. In particular, two-body observables can be calculated analytically in ZREFT. This makes it easier to explore the impact of an S-wave near-threshold resonance on dark matter.

In the absence of electromagnetism, ZREFT is a systematically improvable effective field theory. The improvability is guaranteed by identifying a point in the parameter space in which the S-wave interactions of winos are scale invariant in the low-energy limit, and can therefore be described by an effective field theory that is a renormalization-group fixed point. At the RG fixed point, the mass splitting $\delta$ between the charged wino and the neutral wino is 0, and the corresponding unitarity mass is $M_*=2.22$~TeV. In Ref.~\cite{Braaten:2017gpq}, it was verified explicitly that, in the absence of electromagnetic interactions, ZREFT at NLO provides systematic improvements in the predictions of ZREFT at LO at $\delta = 170$~MeV and the corresponding unitarity mass $M_* = 2.88$~TeV. 
 
In this companion paper to Ref.~\cite{Braaten:2017gpq}, we carried out the Coulomb resummation that is needed to calculate the quantitative predictions of ZREFT at LO. The T-matrix elements for wino-wino scattering are given analytically in eqs.~\eqref{eq:T00LO} and \eqref{eq:T01,11LO}. An analytic equation for the binding energy of the wino-pair bound state is given in eq.~\eqref{eq:gammaLO-eq}. The parameters of ZREFT at LO are the kinematic parameters $M$ and $\delta$ and the interaction parameters $\alpha = 1/137$, $\phi$, and $\gamma_0 = 1/a_0$. The interaction parameters $\phi$ and $\gamma_0$ can be determined by matching predictions of ZREFT at LO for scattering amplitudes with results calculated by solving the Schr\"odinger equation for NREFT numerically. An accurate Pad\'e approximant of $a_0(M)$ for $M$ near the first unitarity mass $M_*=2.39$~TeV is given in eq.~\eqref{eq:a0Pade}. The mixing angle $\phi$ can be determined from NREFT calculations of the effective range $r_0$ by using the matching condition in eq.~\eqref{eq:r0match}. An accurate Pad\'e approximant of $r_0(M)$ for $M$ near $M_*$ is given in eq.~\eqref{eq:r0Pade}. The $M$-dependent mixing angle $\phi(M)$ obtained by matching $r_0(M) $ as a function of $M$ is shown in figure~\ref{fig:tanphivsM}. The mixing angle $\phi(M_*)$ determined by matching $r_0$ at unitarity is given by $\tan \phi(M_*) = 0.877$.

The accuracy of the predictions of ZREFT at LO as functions of the wino mass $M$ was illustrated by the neutral-wino shape parameter $s_0$ and by the binding energy $E_{(ww)}$ of the wino-pair bound state. Accurate predictions away from unitarity require using an $M$-dependent mixing angle $\phi(M)$, such as  that shown in figure~\ref{fig:tanphivsM}. The error in the prediction of $s_0$ remains small thoughout the region of validity of ZREFT, as shown in the right panel of figure~\ref{fig:r0s0vsM}. The error in the prediction of $E_{(ww)}$ increases with $M$, but it also remains small in the region of validity of ZREFT, as shown in  figure~\ref{fig:bindingenergyLO}.

The accuracy of the predictions of ZREFT at LO as functions of the energy $E$ was illustrated by using the wino-wino cross sections at the unitarity mass $M_*=2.39$~TeV. ZREFT at LO gives accurate predictions for the neutral-wino elastic cross section $\sigma_{0 \to 0}$ for $E < 2\delta$ and for $E> 2\delta$, as shown in figure~\ref{fig:sigma00vsE-LO}. Its predictions in the resonance region just below $2 \delta$ have the correct qualitative behavior, as shown in figure~\ref{fig:sigma00vsE-LO_th}. ZREFT at LO gives reasonably good predictions for the charged-to-neutral transition cross section $\sigma_{0 \to 1}$ and the charged-wino elastic cross section $\sigma_{1 \to 1}$, as shown in figure~\ref{fig:sigma01,11vsE-LO}. Its predictions for $\sigma_{1 \to 1}$ in the oscillation region just above $2 \delta$ have the correct qualitative behavior, as shown in figure~\ref{fig:sigma11vsE-LO_th}. More accurate predictions could be obtained by using ZREFT at NLO, which has two additional relevant parameters.

One of the primary motivations for the development of ZREFT for winos was the  calculation of the ``Sommerfeld enhancement'' of the annihilation of a pair of winos into electroweak gauge bosons when the wino mass is near a resonance at the neutral-wino pair threshold. Wino-pair annihilation also affects other aspects of the few-body physics for low-energy winos. For example, the neutral-wino elastic cross section does not actually diverge at a unitarity mass, but it instead has a very narrow peak \cite{Blum:2016nrz}. The effects of wino-pair annihilation on low-energy winos can be taken into account in ZREFT by analytically continuing real parameters to complex values. Since two-body observables for winos can be calculated analytically in ZREFT, the effects of wino-pair annihilation can also be taken into account analytically. The results are presented in another companion paper \cite{BJZ-Annihilation}.

\begin{acknowledgments}
This research project was stimulated by a discussion with M.~Baumgart. We thank S~K\"onig for useful discussions of Coulomb effects in zero-range effective field theory. We thank T.~Slatyer for helpful discussions on the partial wave expansion for the Coulomb interaction. This work was supported in part by the Department of Energy under grant DE-SC0011726.
\end{acknowledgments}

\appendix

\section{Lippmann-Schwinger equation with Coulomb resummation}
\label{app:LSeq}

In the Zero-Range Model with Coulomb resummation, the $2\times2$ matrix $\bm{\mathcal{A}}(E)$ of transition amplitudes can be expressed in the form in eq.~\eqref{eq:A-sC}, where $\mathcal{A}_C(E)$ is the Coulomb amplitude in eq.~\eqref{eq:ACoulomb}, $W_1(E)$ is the amplitude for the creation of $w^+ w^-$ at a point in eq.~\eqref{eq:S-eta}, and $\bm{\mathcal{A}}_s(E)$ is the $2 \times 2$ matrix of short-distance transition amplitudes. In this appendix, we solve the Lippmann-Schwinger integral equations for $\bm{\mathcal{A}}_s(E)$. We also use unitarity to determine the amplitude $W_1(E)$.

\subsection{Short-distance transition amplitudes}

\begin{figure}[t]
\centering
\includegraphics[width=0.9\linewidth]{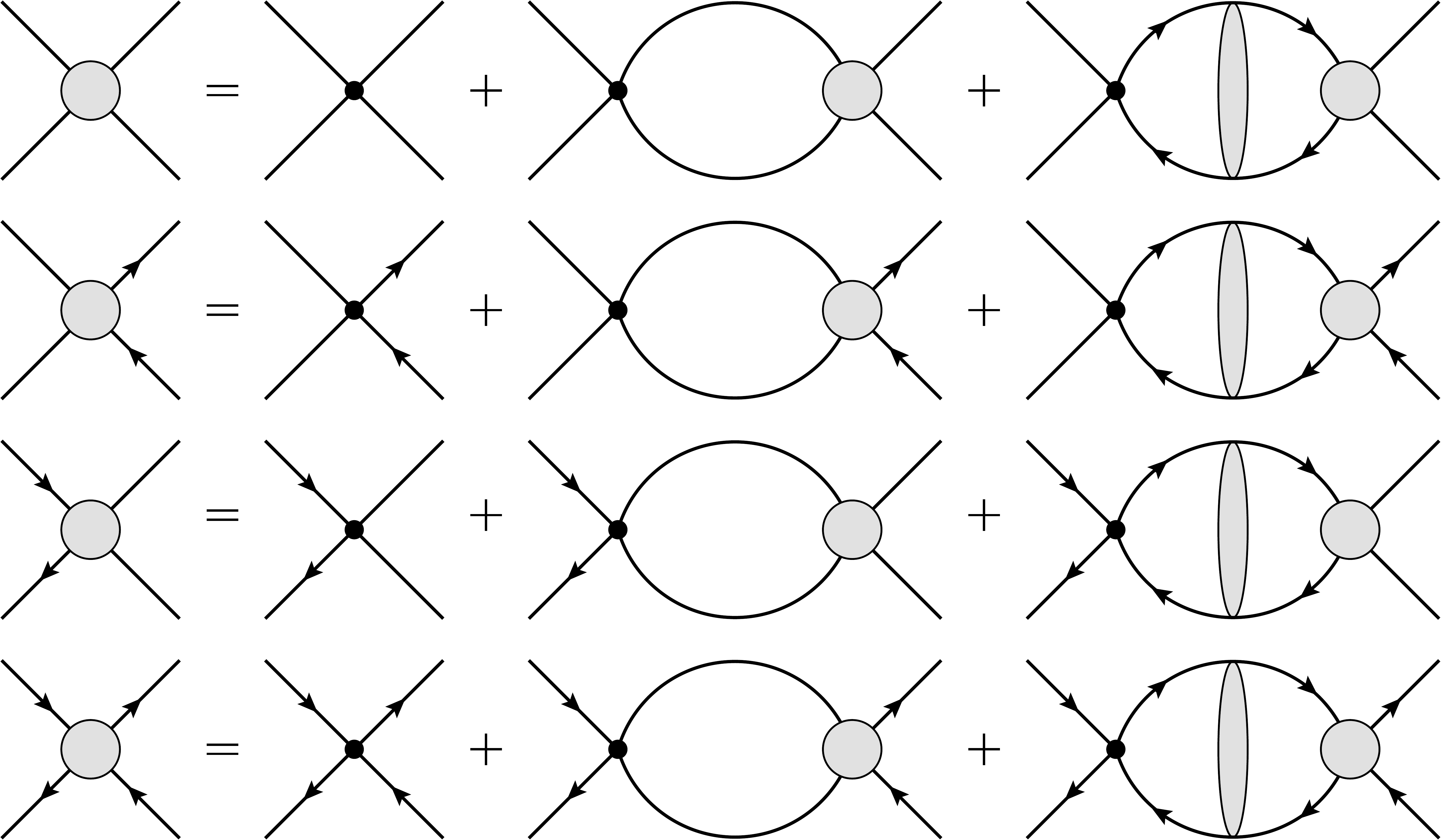}
\caption{Diagrammatic representation of the coupled-channel Lippmann-Schwinger integral equations for the short-distance transition amplitudes ${\cal A}_{s,00}(E)$, ${\cal A}_{s,01}(E)$, ${\cal A}_{s,10}(E)$, and ${\cal A}_{s,11}(E)$. Each of the charged-wino bubbles is the sum of diagrams in figure~\ref{fig:bubbleC}.}
\label{fig:LSeq}
\end{figure}

In the Zero-Range Model discussed in section~\ref{sec:ZRM}, there are two wino-wino channels for which there are zero-range interactions: a pair of neutral winos in the S-wave spin-singlet channel, which we label by 0, and a pair of charged winos in the S-wave spin-singlet channel, which we label by 1. The S-wave spin-singlet transition amplitudes have the same Pauli spinor structure as the zero-range interaction vertices. They can be expressed as $\mathcal{A}_{ij}(E)$ multiplied by the spin-singlet projector $\tfrac12(\delta^{ac}\delta^{bd} - \delta^{ad} \delta^{bc})$, where $i$ and $j$ are the incoming and outgoing channels, $a$ and $b$ are Pauli spinor indices for the incoming lines, and $c$ and $d$ are Pauli spinor indices for the outgoing lines. The transition amplitudes $\mathcal{A}_{ij}(E)$ are functions of the total energy $E$ in the center-of-mass frame. They do not depend separately on the energies and momenta of the incoming and outgoing lines.
 
The Lippmann-Schwinger integral equations for the short-distance transition amplitudes can be expressed as the diagrammatic equations in figure~\ref{fig:LSeq}. In the momentum representation, the  Lippmann-Schwinger equation for the $2 \times 2$ symmetric matrix $\bm{\mathcal{A}}_s(E)$ can be expressed as a matrix equation:
\begin{equation}
\label{eq:LSequation}
\bm{\mathcal{A}}_s(E) = -\bm{\lambda}
+ \bm{\lambda}\;\bm{I}(E)\;\bm{\mathcal{A}}_s(E),
\end{equation}
where $\bm{\lambda}$ is a symmetric matrix of bare coupling constants,
\begin{equation}
\bm{\lambda} = \begin{pmatrix}
\lambda_{00} & \lambda_{01} \\ \lambda_{01} & \lambda_{11}
\end{pmatrix} ,
\label{eq:LOcontactmatrix}
\end{equation}
and $\bm{I}(E)$ is a diagonal matrix of bubble amplitudes:
\begin{equation}
\bm{I}(E) = \begin{pmatrix}
\tfrac12 I_0(E) & 0 \\ 0 & J_1(E)
\end{pmatrix} .
\label{eq:looptmatrix}
\end{equation}
The factor of $\frac12$ in the upper diagonal entry is a symmetry factor. The loop integrals are ultraviolet divergent. They can be regularized using dimensional regularization in $d = 3 - 2 \epsilon$ spatial dimensions. After integrating over the loop energy by contours, the neutral-wino bubble amplitude $I_0(E)$ is
\begin{equation}
I_0(E) = -M \left( \frac{\Lambda}{2} \right)^{3-d} \int \!\! \frac{d^dk}{(2\pi)^d}
\frac{1}{k^2 - ME -i\epsilon},
\label{eq:I0intk}
\end{equation}
where $\Lambda$ is an arbitrary renormalization scale. The integral can be evaluated analytically. The linear ultraviolet divergence in $d=3$ spatial dimensions appears as a pole in $d-2$ with residue $M \Lambda/4\pi$. The integral can be renormalized by {\it power divergence subtraction} \cite{Kaplan:1998tg}, in which the limit $d \to 3$ is taken after subtracting the pole in $d-2$. The  resulting loop integral is 
\begin{equation}
I_0(E) =
- \frac{M}{4\pi}\big[ \Lambda- \kappa_0(E)\big] ,
\label{eq:I0-E}
\end{equation}
where $\kappa_0(E)$ is the  function of the complex energy $E$ defined in eq.~\eqref{eq:kappa0}. The charged-wino bubble amplitude $J_1(E)$ was evaluated analytically using dimensional regularization by Kong and Ravndal\footnote{The function $J_1$ was denoted by $\bar J_0$ in Ref.~\cite{Kong:1999sf}.} \cite{Kong:1999sf}. It can be expressed as the sum of discrete contributions from Coulomb bound states and a dimensionally regularized integral over the relative momentum of scattering states:
\begin{eqnarray}
J_{1}(E) &=& \frac{\alpha^3 M^3}{8 \pi} \sum_{n=1}^{\infty}\frac{1}{n^3(E-E_n)}
\nonumber\\
&& - M \left( \frac{\Lambda}{2} \right)^{3-d} \int \!\! \frac{d^dk}{(2\pi)^d}
\frac{2 \pi \eta(k)}{\exp\! \big(2 \pi \eta(k)\big) - 1}\;
\frac{1}{k^2 - M(E - 2\delta)-i\epsilon},
\end{eqnarray}
where $E_n$ is the energy of the Coulomb bound state in eq.~\eqref{eq:E-n} and $\eta(k) = -\alpha M/(2k)$. The integral has a  linear ultraviolet divergence in $d=3$ spatial dimensions that appears as a pole in $d-2$ with residue $M \Lambda/4\pi$. In the {\it power divergence subtraction} regularization scheme \cite{Kaplan:1998tg}, the linear divergence is canceled by subtracting the pole in $d-2$. The integral also has a logarithmic ultraviolet divergence in $d=3$ spatial dimensions that appears as a pole in $d-3$. After subtracting from the integrand the terms that give the poles in $d-2$ and $d-3$, the remaining integral can be evaluated analytically in $d=3$ dimensions. The final result for the bubble amplitude in the limit $d \to 3$ is
\begin{equation}
J_1(E) = 
- \frac{M}{4\pi} \left[ \Lambda + \alpha M \left(\frac{1}{3-d}  
 +  \log\frac{\sqrt{\pi}\Lambda}{\alpha M} + 1 - \frac32 \gamma\right) - K_1(E)  \right] ,
\label{eq:J1-E}
\end{equation}
where $K_1(E)$ is the function of the complex energy $E$ defined in eq.~\eqref{eq:K1-E} and $\gamma$ is Euler's constant. This result was first calculated by Kong and Ravndal in ref.~\cite{Kong:1999sf}, and it was verified in ref.~\cite{Konig:2015aka}.

To solve the integral equation in eq.~\eqref{eq:LSequation}, we multiply by $\bm{\mathcal{A}}_s^{-1}$ on the right and $\bm{\lambda}^{-1}$ on the left and then rearrange:
\begin{equation}
\bm{\mathcal{A}}_s^{-1}(E) = -\bm{\lambda}^{-1} + \bm{I}(E) .
\label{eq:LSsolution}
\end{equation}
The dependence of the amplitudes $\mathcal{A}_{s,ij}(E)$ on the renormalization scale can be eliminated by choosing the bare parameters $\lambda_{ij}$ to depend on $\Lambda$ in such a way that
\begin{eqnarray}
\bm{\lambda}^{-1} &=& 
\frac{M}{8 \pi} 
\begin{pmatrix} \gamma_{00}  &\sqrt{2}\, \gamma_{01} \\ \sqrt{2}\,\gamma_{01} & 2\gamma_{11}  \end{pmatrix} 
- \frac{M\Lambda}{8\pi} \begin{pmatrix} ~1~  & ~0~ \\ 0 & 2  \end{pmatrix}
\nonumber \\
&& 
 - \frac{\alpha M^2}{4\pi} \left(\frac{1}{3-d}  
+ \log\frac{\sqrt{\pi}\Lambda}{\alpha M} + 1 - \frac32 \gamma  \right)
 \begin{pmatrix} ~0~  & ~0~ \\ 0 & 1  \end{pmatrix}.
\label{eq:lambda-gamma}
\end{eqnarray}
This defines physical scattering parameters $\gamma_{00}$, $\gamma_{01}$, and $\gamma_{11}$ with dimensions of momentum. Substituting these relations into eq.~\eqref{eq:LSsolution}, we have
\begin{equation}
\label{eq:Ainverse2}
\bm{\mathcal{A}}_s^{-1}(E) = \frac{M}{8\pi}
\begin{pmatrix} -\gamma_{00} + \kappa_0(E)  & -\sqrt{2}\,\gamma_{01} \\ 
 -\sqrt{2}\,\gamma_{01} & 2\big[- \gamma_{11} +K_1(E) \big] 
\end{pmatrix} .
\end{equation}
The inverse $\bm{\mathcal{A}}_s(E)$ of this $2 \times 2$ matrix is the matrix of short-distance transition amplitudes for the two coupled channels.

\subsection{Amplitude for creating a charged-wino pair at a point}

To complete the calculation of the transition amplitudes for the Zero-Range Model with Coulomb resummation, we must determine the amplitude $W_1(E)$ for creating $w^+ w^-$ at a point with total energy $E$. We use the unitarity condition for $\bm{\mathcal{A}}(E)$ and the optical theorem for the charged-wino bubble amplitude $J_1(E)$.

The unitarity condition for $\bm{\mathcal{A}}(E)$ in eq.~\eqref{eq:A-unitarity} can be reduced to a similar equation for the short-distance amplitudes:
\begin{eqnarray}
\bm{\mathcal{A}}_s(E) - \bm{\mathcal{A}}_s(E)^* &=& 
- \frac{1}{8 \pi}\bm{\mathcal{A}}_s(E)
\begin{pmatrix} ~1~  & 0  \\  0  & W_1(E) \end{pmatrix} 
\nonumber\\
&&\times
 \bm{M}^{1/2} \big[ \bm{\kappa}(E)  - \bm{\kappa}(E)^* \big]  \bm{M}^{1/2} 
\begin{pmatrix} ~1~  & 0  \\  0  & W_1(E)^* \end{pmatrix} 
\bm{\mathcal{A}}_s(E)^*,
\label{eq:As-unitarity}
\end{eqnarray}
provided $W_1(E)$ at a real energy $E$ satisfies the identity
\begin{equation}
W_1(E) - W_1(E)^* =
- \frac{M}{4\pi} \mathcal{A}_C(E) \big[ \kappa_1(E)  - \kappa_1(E)^* \big] W_1(E)^*.
\label{eq:discW1}
\end{equation}
Using  the explicit expression for the Coulomb amplitude in eq.~\eqref{eq:ACoulomb}, this implies that  for real energies $E>2\delta$, $W_1$ must satisfy
\begin{equation}
W_1(E)/W_1(E)^* = \Gamma(1+i \eta)/\Gamma(1-i \eta),
\label{eq:S/S*}
\end{equation}
where $\eta$ is the function of $E$ defined in eq.~\eqref{eq:eta-def}.

\begin{figure}[t]
\centering
\includegraphics[width=0.6\linewidth]{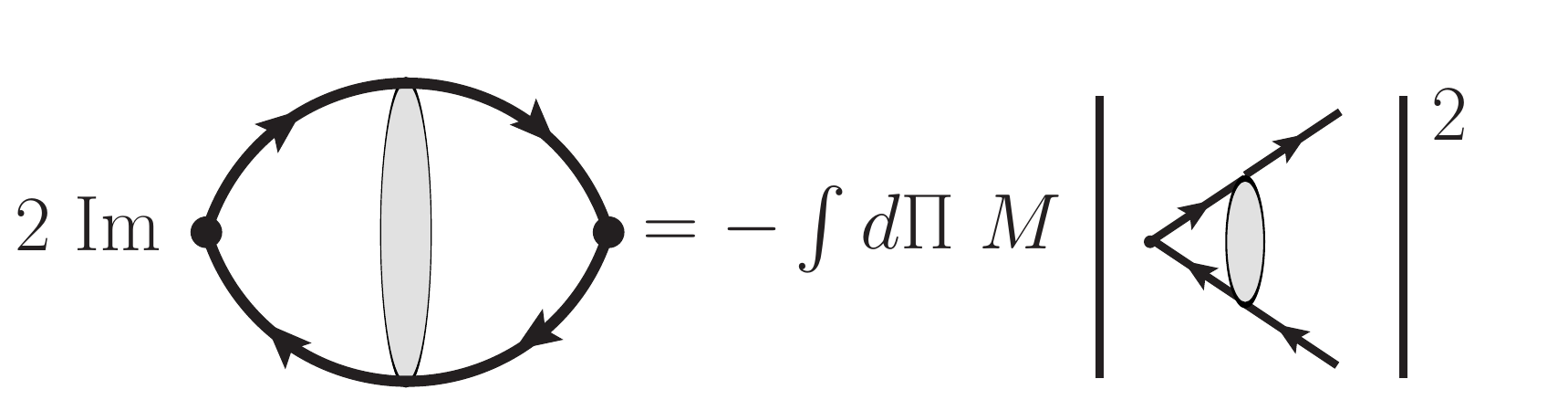}
\caption{The discontinuity in the bubble amplitude $J_1(E)$ at a real energy $E > 2 \delta$ is proportional to the absolute square of the amplitude $W_1(E)$ for creating $w^+ w^-$ at a point.}
\label{fig:ImJ1}
\end{figure}

The optical theorem for the charged-wino bubble amplitude is represented diagrammatically in figure~\ref{fig:ImJ1}. It determines the imaginary part of the function $J_1(E)$ at real energies $E > 2 \delta$:
\begin{equation}
J_1(E) - J_1(E)^* = -i \big| W_1(E) \big|^2 \frac{M \sqrt{M(E- 2 \delta)}}{2 \pi}.
\label{eq:ImJ1}
\end{equation}
The last factor is the phase space integral for $w^+ w^-$ with total energy $E$. The function $J_1(E)$ is given explicitly in eq.~\eqref{eq:J1-E}. Its imaginary part comes from the imaginary part of the function $K_1(E)$ defined in eq.~\eqref{eq:K1-E}. For real values of $E$, that function  satisfies the identity
\begin{equation}
K_1(E) - K_1(E)^*  = C^2(E) \big[ \kappa_1(E)  - \kappa_1(E)^* \big],
\label{eq:ImK1-E}
\end{equation}
where  $C^2$ is the Sommerfeld factor in eq.~\eqref{eq:Sommerfeld}. For $E<  2 \delta$, both sides are 0. For $E> 2 \delta$, the identity follows from a property of the function $\psi(z)$:
\begin{equation}
\psi(z) - \psi(-z)  = - \frac{1}{z} - \frac{\pi}{\tan(\pi z)}.
\label{eq:psi-z}
\end{equation}
Comparing eqs.~\eqref{eq:ImJ1} and \eqref{eq:ImK1-E}, we find that for real energies $E>2\delta$, $W_1(E)$ must satisfy
\begin{equation}
W_1(E)\, W_1(E)^* = C^2(E).
\label{eq:SS*-C}
\end{equation}
Combining eqs.~\eqref{eq:S/S*} and \eqref{eq:SS*-C}, we obtain the amplitude $W_1(E)$ for creating $w^+ w^-$ at a point in eq.~\eqref{eq:S-eta}. This expression was derived for real $E> 2 \delta$, but it can be extended to complex $E$ by analytic continuation.

On the right side of eq.~\eqref{eq:As-unitarity}, all the diagonal matrices between $\bm{\mathcal{A}}_s$ and $\bm{\mathcal{A}}_s^*$ commute. Since the product of $\text{diag}(1,W_1)$ and $\text{diag}(1,W_1^*)$ is $\text{diag}(1,C^2)$, they can both be replaced by $\text{diag}(1,C)$. By multiplying  eq.~\eqref{eq:As-unitarity} by a prefactor of $\text{diag}(1,C)$ and  by a postfactor of $\text{diag}(1,C)$, we find that the matrix $\text{diag}(1,C) \bm{\mathcal{A}}_s \text{diag}(1,C)$ satisfies the unitarity condition in eq.~\eqref{eq:A-unitarity}.


\end{document}